\documentclass[a4paper,11pt]{article}
\usepackage{pos}
\usepackage{colortbl,tabularx}
\usepackage{placeins}

\definecolor{darkblue}{rgb}{0.15,0.15,0.6}
\definecolor{darkgreen}{rgb}{0.0,0.4,0.0}
\definecolor{lightgreen}{rgb}{0.0,0.7,0.0}
\definecolor{citecolor}{rgb}{0.0,0.3,0.0}
\definecolor{darkorange}{rgb}{0.811, 0.408, 0}
\definecolor{darkyellow}{rgb}{0.75, 0.6, 0}
\definecolor{lightorange}{rgb}{1, 0.8, 0.6}
\definecolor{lightblue}{HTML}{aee4f2}

\newcommand{\wm}{\phantom{-}}
\newcommand{\bss}[1]{\ensuremath{{\boldsymbol{#1}}}}
\newcommand{\red}[1]{{\textcolor{red}{#1}}}

\newcommand{\magen}[1]{{\textcolor{magenta}{#1}}}

\title{Quark flavor physics with lattice QCD}

\author*{Stefan Meinel}

\affiliation{Department of Physics, University of Arizona, Tucson, AZ 85721, USA}

\emailAdd{smeinel@arizona.edu}

\abstract{This is an overview of quark flavor physics as presented in a plenary talk at Lattice 2023. In the first part, I discuss the main processes and lattice-QCD inputs used to determine the Wolfenstein parameters of the Cabibbo-Kobayashi-Maskawa matrix. In the second part, I review selected further processes that are being used to search for physics beyond the Standard Model. New results presented at Lattice 2023 are referenced throughout, but detailed discussions are limited to selected work published prior to the conference. QED corrections, inclusive decays on the lattice, and processes involving hadronic resonances are not discussed in detail here, as they were covered in other plenary talks at Lattice 2023 and Lattice 2022.
}

\FullConference{The 40th International Symposium on Lattice Field Theory (Lattice 2023)\\
July 31st - August 4th, 2023\\
Fermi National Accelerator Laboratory\\}

\begin{document}
\maketitle

\section{Introduction}

\noindent The Standard Model of particle physics has the fascinating property that each of the five different fermion representations of the gauge group $SU(3)\times SU(2)\times U(1)$ comes in three copies, the \textit{generations} \cite{Cabibbo:1963yz,Bjorken:1964gz,Glashow:1970gm,Kobayashi:1973fv,Harari:1977kv}:
\begin{eqnarray}
\nonumber && \hspace{4ex}1\hspace{7ex}2\hspace{7ex}3 \\
\nonumber  Q^\prime_L&=&\left( \left(\begin{array}{c} u^\prime_L \\ d^\prime_L \end{array}\right), \left(\begin{array}{c} c^\prime_L \\ s^\prime_L \end{array}\right), \left(\begin{array}{c} t^\prime_L \\ b^\prime_L \end{array}\right) \right), \\ 
\nonumber  U^\prime_R&=&\left( \hspace{2.5ex}  u^\prime_R ,  \hspace{4.5ex} c^\prime_R, \hspace{4.5ex} t^\prime_R \hspace{2ex} \right), \\
\nonumber  D^\prime_R&=&\left(   \hspace{2.5ex} d^\prime_R ,  \hspace{4.5ex} s^\prime_R, \hspace{4.5ex} b^\prime_R \hspace{1.5ex} \right), \\
\nonumber  L^\prime_L &=&\left(\! \left(\begin{array}{c} {\nu_e}^\prime_L \\ e^\prime_L \end{array} \!\right), \left(\!\begin{array}{c} {\nu_\mu}^\prime_L \\ \mu^\prime_L \end{array}\!\right), \left(\!\begin{array}{c} {\nu_\tau}^\prime_L \\ \tau^\prime_L \end{array}\!\right) \right), \\
 E^\prime_R &=& \left(  \hspace{2.5ex} e^\prime_R , \hspace{4.5ex} \mu^\prime_R, \hspace{4.5ex} \tau^\prime_R \hspace{2ex} \right).
\end{eqnarray} 
The species labels $u$, $c$, $t$, $d$, $s$, $b$, $e$, $\mu$, $\tau$ are called \textit{flavors}. The $^\prime$ indicates the use of the gauge basis. In the absence of flavor-violating interactions, we would have a $U(3)^5$ global flavor symmetry. In the Standard Model, the only origin of flavor symmetry violation (and CP violation) is the Yukawa interaction of the fermions with the Higgs field $\phi$ \cite{Weinberg:1967tq,Kobayashi:1973fv}:
\begin{equation}
\mathcal{L}_{\rm Yukawa}\:=\:  - \overline{Q^\prime_{Li}} Y^U_{ij} U^\prime_{Rj} \tilde{\phi} - \overline{Q^\prime_{Li}} Y^D_{ij} D^\prime_{Rj} \phi -\overline{L^\prime_{Li}} Y^E_{ij} E^\prime_{Rj} \phi \:+\: {\rm h.c.}\,.
\end{equation}
When $\phi$ acquires its vacuum expectation value $\langle \phi \rangle = (0, v/\sqrt{2})$, these couplings produce the fermion mass terms. In the quark sector, the unitary field transformations that diagonalize the mass matrices,
\begin{equation}
\begin{array}{ll}
 U^\prime_L = V^U_L U_L, & U^\prime_R = V^U_R U_R, \\
 D^\prime_L = V^D_L D_L, & D^\prime_R = V^D_R D_R, \\
\end{array}
\end{equation}
do not cancel in the charged current coupling to the $W$ field,
\begin{eqnarray}
 \mathcal{L}_{\rm c.c.}\:\:=\:\:-\frac{g}{\sqrt{2}}  \overline{U^\prime_{Li}}\gamma^\mu D^\prime_{Li}\:W^+_\mu \:+\: {\rm h.c.} &=& -\frac{g}{\sqrt{2}}  \overline{U_{Li}} \underbrace{(V^{U\dag}_L V^D_L)_{ij}}_{= \displaystyle V_{ij}}  \gamma^\mu D_{Lj}\:W^+_\mu  \:+\: {\rm h.c.},
\end{eqnarray}
giving rise to the unitary Cabibbo-Kobayashi-Maskawa (CKM) quark mixing matrix
\begin{equation}
V =  \left(\! \begin{array}{lll} V_{ud}  & V_{us} & V_{ub} \\ V_{cd} & V_{cs} & V_{cb} \\ V_{td} & V_{ts} & V_{tb} \end{array}\!  \right). 
\end{equation}
After eliminating unobservable phase factors, $V$ can be written in terms of four parameters.

\vspace{2ex}

\noindent Some of the fundamental questions in flavor physics are:
\begin{itemize}\setlength{\itemsep}{0ex}
 \item What is the origin of the three generations?
 \item What is the origin of the hierarchies in the fermion masses and mixing matrices?
 \item Are there other sources of flavor-violating interactions and CP violation beyond the Standard Model?
\end{itemize}
In most of the more fundamental theories that have been proposed to address the deficiencies of the Standard Model, the answer to the third question is ``yes''. The precision study of flavor-changing processes is therefore a powerful tool for discovering new physics. Lattice-QCD calculations play an essential role in this effort, as discussed in the following.

\section{Main processes used to determine the CKM parameters}

\noindent An insightful parametrization of the CKM matrix was proposed by Lincoln Wolfenstein 40 years before this Lattice conference \cite{Wolfenstein:1983yz}. This parametrization is based on the observation that $V$ is close to the identity matrix. Among the off-diagonal elements, $V_{us}$ was already well determined to be close to $0.22$, and Wolfenstein estimated $V_{cb}\approx0.06$ from the $B$ meson lifetime that had just been measured at the PEP storage ring \cite{Lockyer:1983ev}. This hierarchy suggested an expansion in powers of a parameter $\lambda$, where $V_{us}$ is of first order and $V_{cb}$ is of second order. Writing $V_{us}=\lambda$ and $V_{cb}=A\lambda^2$ and imposing unitarity gives the two-parameter approximation
\begin{equation}
 V \:=\:  \left(\! \begin{array}{ccc} 1-\frac12\lambda^2  & \lambda & 0 \\ -\lambda & 1-\frac12\lambda^2 & A\lambda^2 \\ 0 & -A\lambda^2 & 1 \end{array}\!  \right)\: +\: \mathcal{O}(\lambda^3).
\end{equation}
At higher orders, two additional parameters, denoted as $\rho$ and $\eta$, are needed. Imposing unitarity yields the form
\begin{equation}
 V \:=\:  \left(\! \begin{array}{ccc} 1-\frac12\lambda^2  & \lambda & A\lambda^3(\rho-i\eta) \\ -\lambda & 1-\frac12\lambda^2 & A\lambda^2 \\ A\lambda^3(1-\rho-i\eta) & -A\lambda^2 & 1 \end{array}\!  \right)\: +\: \mathcal{O}(\lambda^4).
\end{equation}
In 1983, only upper limits on $\rho$ and $\eta$ were available \cite{Wolfenstein:1983yz}, and there has been a lot of progress since then. In the next few subsections, I will discuss the current status of determinations of the CKM parameters and the relevant inputs from lattice QCD, following the hierarchy encoded in the Wolfenstein parametrization.

\subsection{Determination of $\bss{\lambda}$}

\noindent The exact definition of this Wolfenstein parameter is $\lambda = \frac{|V_{us}|}{\sqrt{|V_{ud}|^2+|V_{us}|^2}}$ \cite{PDGCKM}. Assuming the CKM unitarity relation $|V_{ud}|^2+|V_{us}|^2+|V_{ub}|^2=1$ and neglecting the very small $|V_{ub}|^2$, a determination of $|V_{ud}|$ alone already gives us $|V_{us}|$ as well, and hence $\lambda$. The most precise direct result for $|V_{ud}|$ comes from the study of superallowed $0^+\to 0^+$ nuclear $\beta$ decays, which are pure vector transitions and therefore fairly insensitive to nuclear/nucleon structure, yielding \cite{PDGVudVus}
\begin{equation}
 |V_{ud}| = 0.97373(11)_{\rm exp.}(9)_{\rm RC}(27)_{\rm NS}.
\end{equation}
This result alone would give $\lambda = 0.2277(13)$. But is the unitarity relation actually satisfied?

\begin{figure}
\begin{center}
\includegraphics[width=0.4\linewidth]{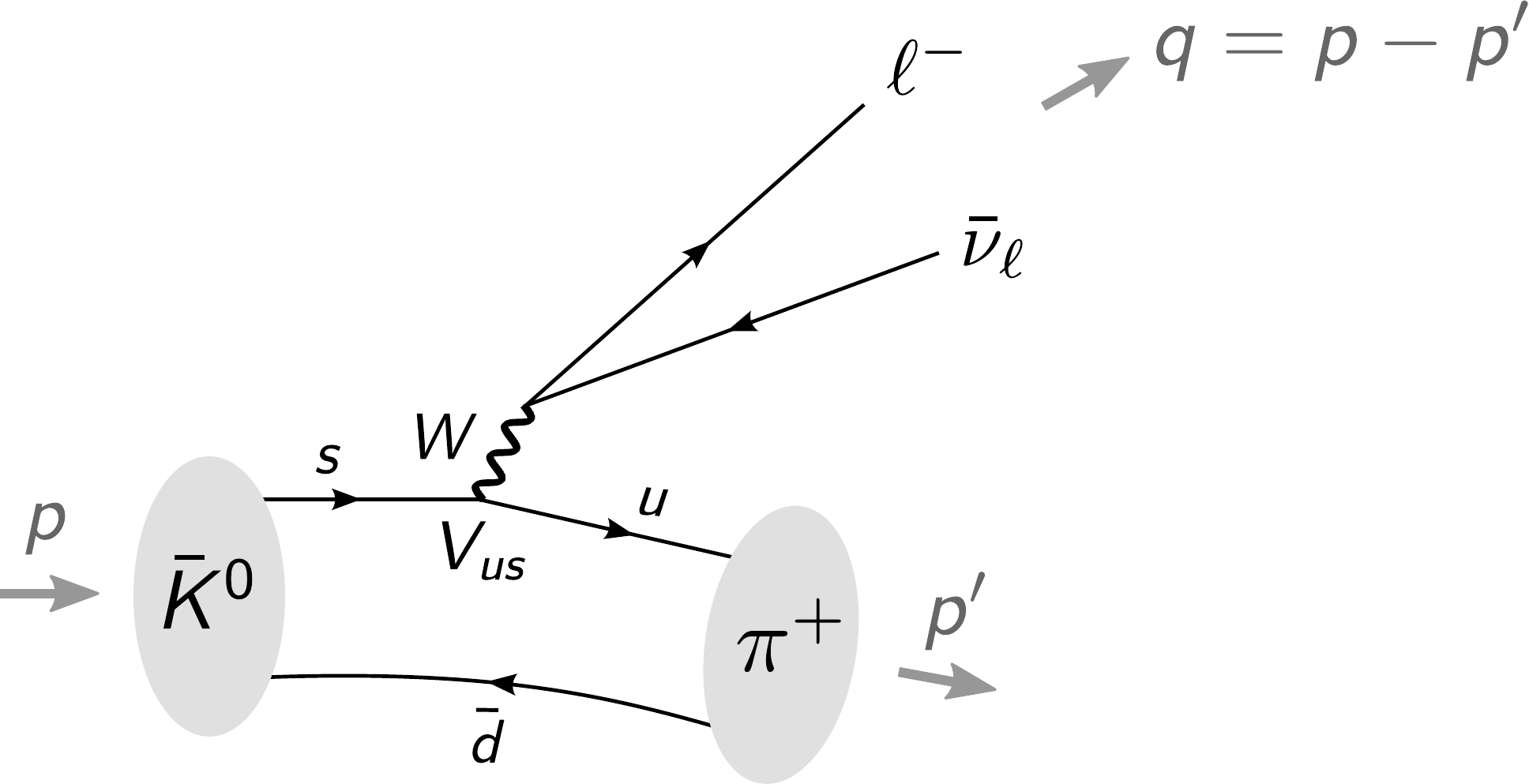}
\end{center}
\caption{\label{fig:kaonSL}A schematic diagram of kaon semileptonic decay in the Standard Model.}
\end{figure}

\begin{figure}
\includegraphics[width=0.49\linewidth]{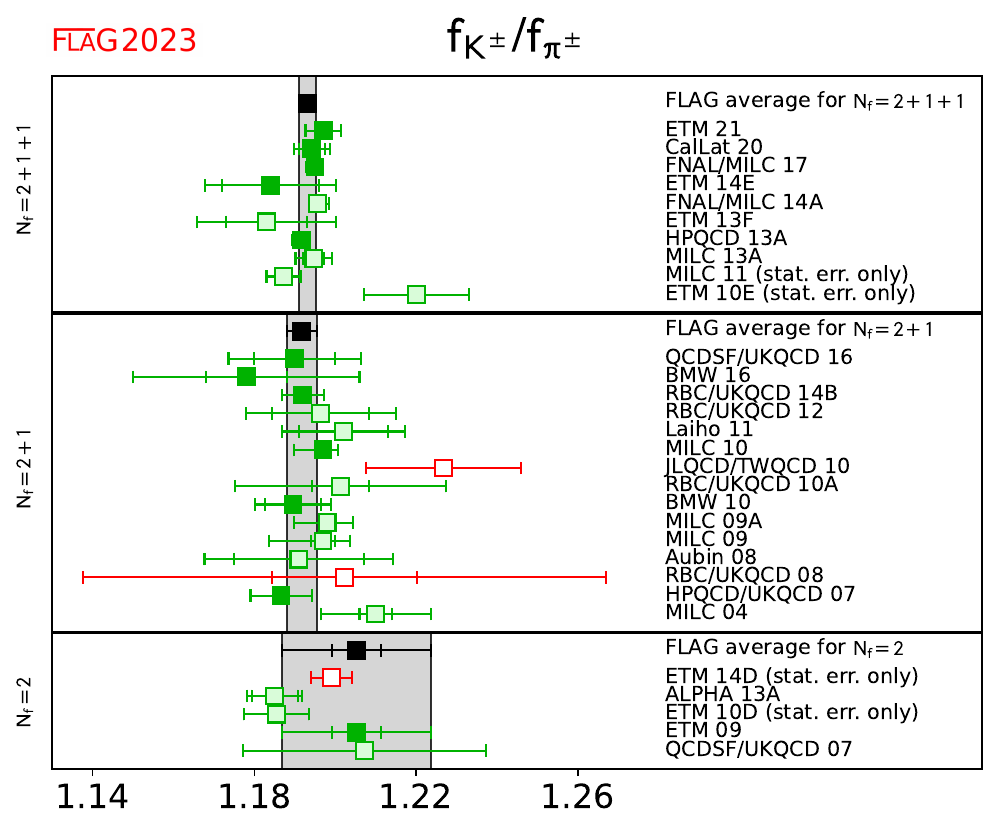} \hfill  \includegraphics[width=0.49\linewidth]{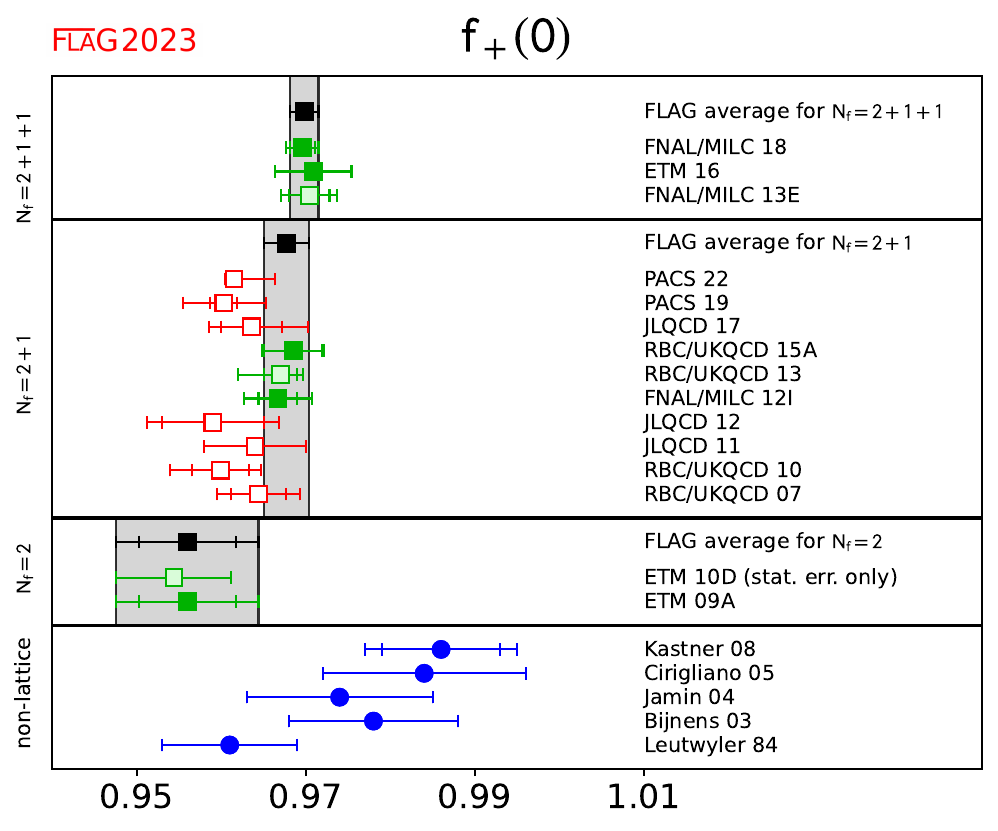}
\caption{\label{fig:RfKfpif+0}Summary of lattice results for the ratio of kaon and pion decay constants (left), and of both lattice and non-lattice results for the kaon semileptonic form factor at $q^2=0$ (right) by FLAG \cite{FlavourLatticeAveragingGroupFLAG:2021npn,FLAGweb}.}
\end{figure}

\noindent We also have the following experimental results for kaon decays \cite{PDGVudVus}:
\begin{eqnarray}
 \displaystyle\frac{\Gamma(K^\pm \to \mu^\pm \nu [\gamma])}{\Gamma(\pi^\pm \to \mu^\pm \nu [\gamma])} \!\!&=&\!\! 1.3367(28), \\
\nonumber && \\
 \frac{\mathrm{d}\Gamma}{\mathrm{d}q^2}(K \to \pi \ell \nu[\gamma]) \hspace{2ex}\underset{\text{non-lattice theory}}{\Rightarrow} \hspace{2ex} f_+(K \to \pi , q^2=0)|V_{us}| \!\!&=&\!\! 0.21635(38)(3).
\end{eqnarray}
Here, $q$ is the 4-momentum transfer, defined as shown in Fig.~\ref{fig:kaonSL}. To get $|V_{us}/V_{ud}|$ and $|V_{us}|$ from these results, we need lattice-QCD calculations of the ratio of decay constants
$f_{K^\pm}/f_{\pi^\pm}$
and of the form factor
$f_+(K \to \pi , q^2=0)$; these quantities parametrize the QCD matrix elements
\begin{eqnarray}
 \langle 0 |\: \bar{u}\gamma^\mu\gamma_5 d \:{|\pi^-(p)\rangle} &=& i p^\mu {f_{\pi^-}}, \\
 \langle 0 |\: \bar{u}\gamma^\mu\gamma_5 s \:{|K^-(p)\rangle} &=& i p^\mu {f_{K^-}}, \\
\nonumber {\langle \pi^+(p^\prime)} |\: \bar{u}\gamma^\mu s \:{|\bar{K}^0(p)\rangle} &=&\left[ (p+p^\prime)^\mu - \frac{m_K^2-m_\pi^2}{q^2}q^\mu \right] {f_+(K \to \pi , q^2)} \\
 &&+         \frac{m_K^2-m_\pi^2}{q^2} q^\mu {f_0(K \to \pi, q^2)}.
\end{eqnarray}
As shown in Fig.~\ref{fig:RfKfpif+0}, lattice calculations of $f_{K^\pm}/f_{\pi^\pm}$ and $f_+(K \to \pi , q^2=0)$ have reached impressive precision, resulting in Flavour Lattice Averaging Group (FLAG) averages with uncertainties of approximately 0.2\% \cite{FlavourLatticeAveragingGroupFLAG:2021npn,FLAGweb}. Note that the decay rates measured by the experiments are fully photon-inclusive. These QED corrections can be calculated in chiral perturbation theory or on the lattice, as done for the leptonic decays in Ref.~\cite{DiCarlo:2019thl}. Using the latter, the FLAG determination of $(|V_{us}|,|V_{ud}|)$ from lattice calculations with $N_f=2+1+1$ dynamical quark flavors, shown as the red ellipse in Fig.~\ref{fig:RfKfpif+0}, is in significant tension with CKM unitarity and with the result for $|V_{ud}|$ from nuclear beta decays. It is interesting to note that there are viable new-physics models that can resolve these tensions; for example, TeV-scale vector-like quarks can introduce small right-handed couplings that will do the job and can also explain the $W$-boson-mass anomaly \cite{Belfatto:2023tbv}.

At this conference, Takeshi Yamazaki presented a new preliminary result for $f_+(K \to \pi , q^2=0)$ from calculations on impressively large lattices of size greater than $(10\text{ fm})^4$ for three different lattice spacings and with physical pion masses \cite{Yamazaki:2023swq}. The corresponding value for $|V_{us}|$ is shown in yellow in Fig.~\ref{fig:RfKfpif+0} and would bring the $(|V_{us}|,|V_{ud}|)$ combination into better agreement with unitarity and with $|V_{ud}|$ from nuclear beta decays. Another possible way to determine $|V_{us}|$ and $|V_{ud}|$ is though inclusive hadronic $\tau$ decays. Antonio Evangelista presented the first fully nonperturbative lattice calculation of the inclusive $\tau$ decay rate in the $\bar{u}d$ flavor channel using spectral reconstruction \cite{Evangelista:2023fmt}, resulting in a $|V_{ud}|$ value with 0.4\% uncertainty that is also shown in Fig.~\ref{fig:RfKfpif+0}. While this result is not competitive with the extremely precise determination from nuclear beta decays, the method can also be applied in the $\bar{u}s$ flavor channel where it will likely be competitive with the meson decays. Further work related to $|V_{ud}|$ and $|V_{us}|$ shown at this conference includes a calculation of $f_\pi$ and $f_K$ presented by Zack Hall \cite{Hall} and investigations of QED corrections presented by Norman Christ \cite{Christ}, Antonin Portelli \cite{Portelli}, Nils Hermansson Truedsson \cite{DiCarlo:2023rlz}, and Jun-sik Yoo \cite{Yoo:2023gln}. A separate plenary talk, given by Matteo Di Carlo, was devoted to the topic of QED corrections \cite{DiCarlo}.

\begin{figure} 
\begin{center}
\includegraphics[width=0.9\linewidth]{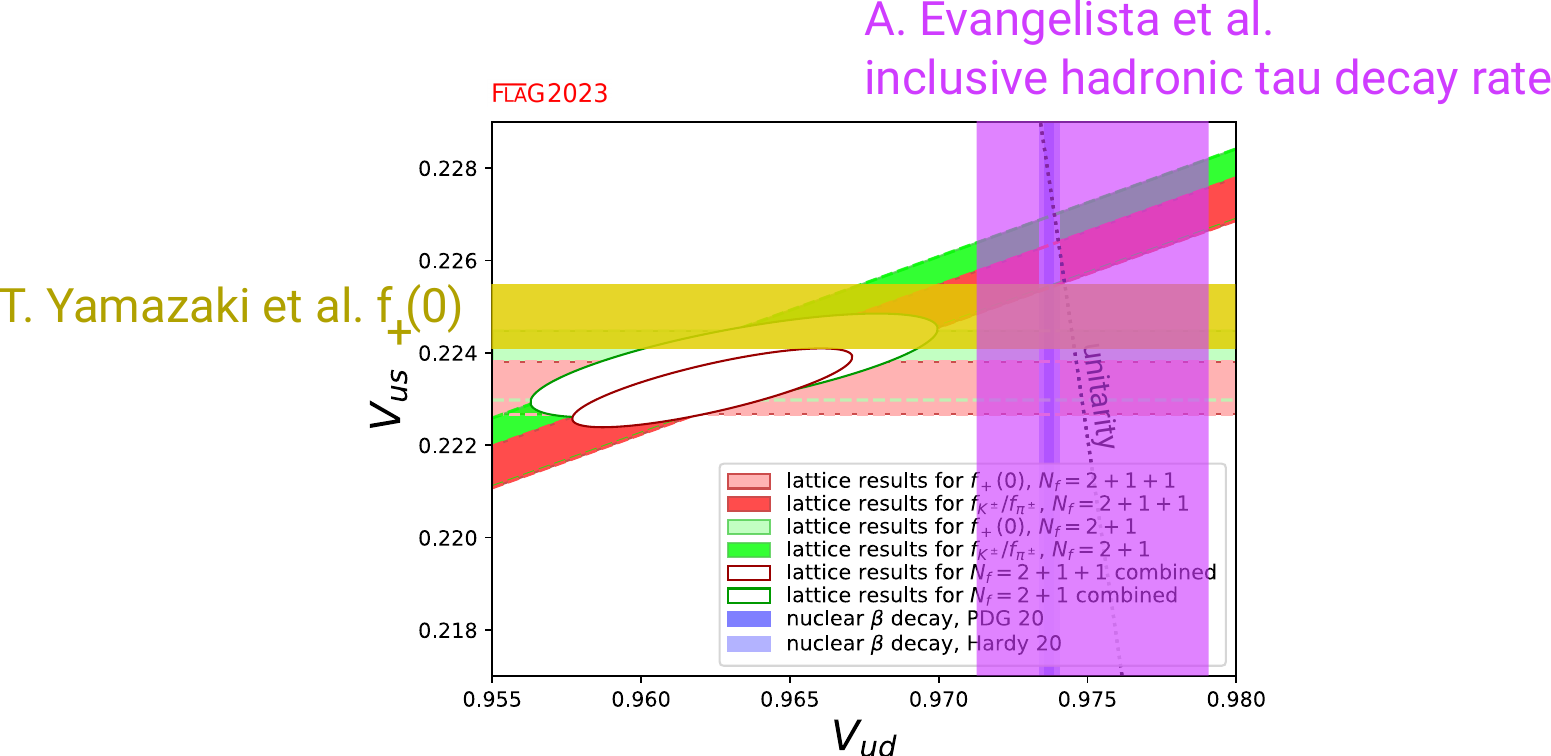}
\end{center}
\caption{Constraints in the $|V_{us}|$-$|V_{ud}|$ plane from leptonic and semileptonic meson decays (red, green, and yellow), nuclear $\beta$ decays (blue), and inclusive hadronic $\tau$ decays \cite{FlavourLatticeAveragingGroupFLAG:2021npn,FLAGweb,Yamazaki:2023swq,Evangelista:2023fmt}.}
\end{figure}

\subsection{Determination of $\bss{A}$}
\label{sec:A}

\noindent The exact definition of the Wolfenstein parameter $A$ is given by $A\lambda^2 = \frac{\lambda}{|V_{us}|} |V_{cb}|$ \cite{PDGCKM}. The next task is therefore to determine $|V_{cb}|$. The most important processes currently used to determine $|V_{cb}|$ are semileptonic $B_{(s)}$ meson decays:
\begin{itemize}\setlength{\itemsep}{0ex}
 \item Inclusive $B\to X_c \ell\nu$ ($\ell=e,\mu$; BaBar, Belle, Belle II, and older experiments),
 \item Exclusive $B\to D\ell\nu$ ($\ell=e,\mu$; BaBar, Belle, Belle II, and older experiments),
 \item Exclusive $B\to D^*\ell\nu$ ($\ell=e,\mu$; BaBar, Belle, Belle II, and older experiments),
 \item Exclusive $B_s\to D_s\mu\nu$ (LCHb),
 \item Exclusive $B_s\to D^*_s\mu\nu$ (LCHb).
\end{itemize}
The exclusive determinations use form factors from lattice QCD. The most precise inclusive determinations use the heavy-quark/operator-product expansion in powers of $1/m_b$ and $\alpha_s$, where hadronic matrix elements of $\Delta B=0$ matrix elements are fitted to experimental data \cite{Gambino:2020jvv}; these calculations use lattice input for $m_b$, $m_c$, $\alpha_s$. There is also substantial progress with direct lattice calculations of inclusive rates. This was covered thoroughly in the Lattice 2022 plenary talks by Takashi Kaneko \cite{Kaneko:2023kxx} and John Bulava \cite{Bulava:2023mjc}.

\begin{table}
\begin{center}
\begin{tabular}{|lccc|}
\rowcolor{gray!30} \hline
                         &  Fermilab/MILC \cite{FermilabLattice:2021cdg}            &  HPQCD \cite{Harrison:2023dzh}               & JLQCD \cite{Aoki:2023qpa}             \\
\hline
\rowcolor{gray!0} { \phantom{ $\int_a^b$} \hspace{-5ex} $u,d,s,(c)$-quark action    \hspace{-2.5ex}}  &  AsqTad (2+1)            &  HISQ (2+1+1)          & domain wall (2+1)            \\
\rowcolor{gray!0} { \phantom{ $\int_a^b$} \hspace{-5ex} $b$-quark action        \hspace{-2.5ex}}  &  Fermilab clover     &  HISQ                & domain wall           \\
\rowcolor{gray!0} { \phantom{ $\int_a^b$} \hspace{-5ex} $B$-meson mass         \hspace{-2.5ex}}  &  $m_{\rm kin}\approx m_{\rm phys}$    &  $m \lesssim 0.93\: m_{\rm phys}$                &    $m \lesssim 0.74\: m_{\rm phys}$        \\
\rowcolor{gray!0} { \phantom{ $\int_a^b$} \hspace{-5ex} $m_\pi$ (MeV)           \hspace{-2.5ex}}  & 180 - 560$^*$            &  135 - 329$^*$            &  230 - 500           \\
\rowcolor{gray!0} { \phantom{ $\int_a^b$} \hspace{-5ex} $a$ (fm)                \hspace{-2.5ex}}  & 0.045 - 0.15         &  0.044 - 0.090          &  0.044 - 0.080         \\
\rowcolor{gray!0} { \phantom{ $\int_a^b$} \hspace{-5ex} $\#$(source-sink separations)                \hspace{-2.5ex}} & 2 ($T$, $T\!+\!1$)  &  3  &  4 \\
\hline
\end{tabular}

\vspace{0.5ex}

{\small  $^*$These are the masses of the lightest pion (taste $\gamma_5$)}
\end{center}

\caption{\label{tab:BDstar}Lattice calculations of the $B\to D^*$ form factors at nonzero recoil.}
\end{table}

\begin{figure}
\begin{center} 
\includegraphics[width=0.6\linewidth]{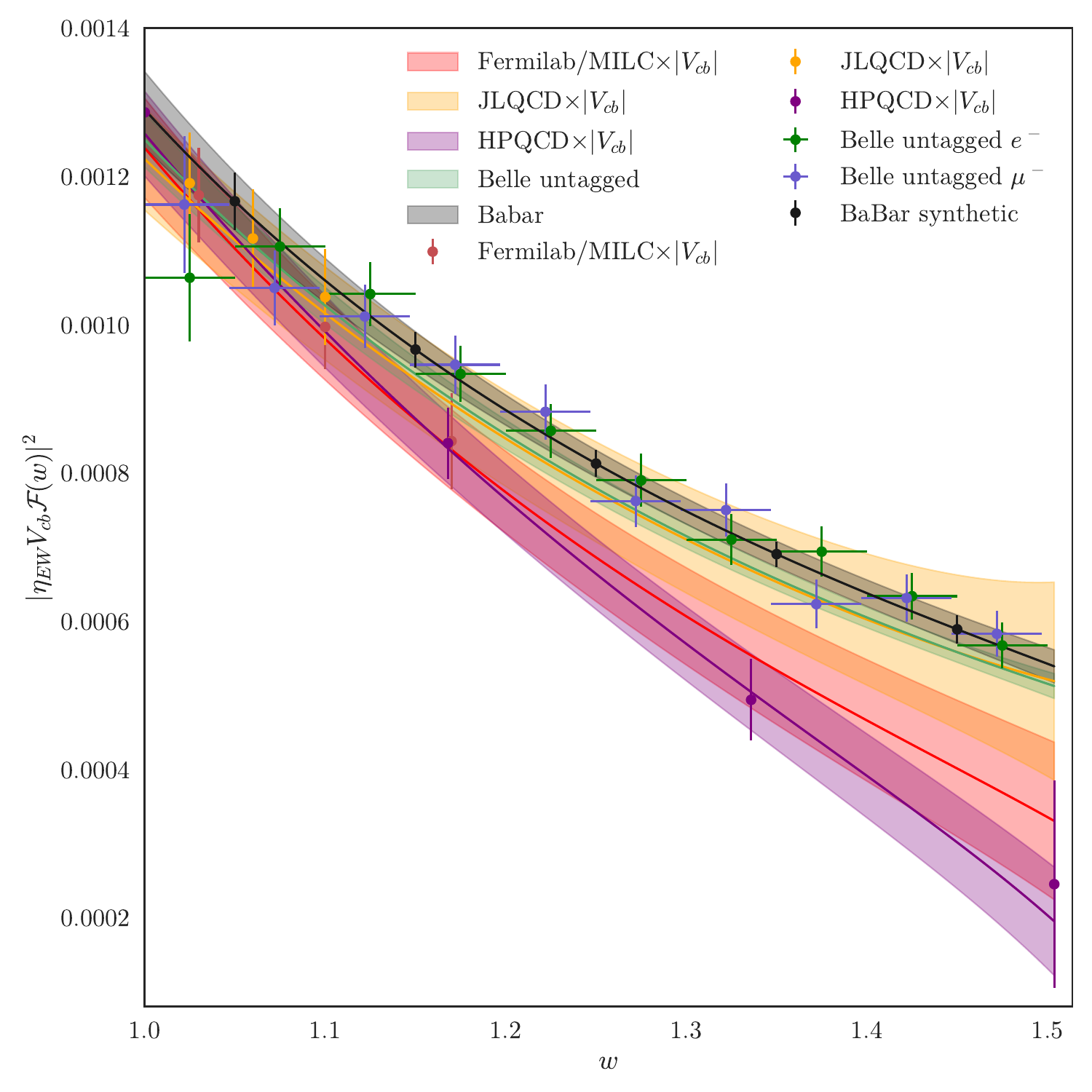}
\end{center}
\caption{\label{fig:BDstarF}Comparison of lattice predictions \cite{FermilabLattice:2021cdg,Harrison:2023dzh,Aoki:2023qpa} and experimental results \cite{BaBar:2019vpl,Belle:2018ezy} for the combination of form factors that appears in the $B\to D^*\ell\nu$ differential decay rate. The black and green curves are from BGL fits to the experimental data  [plot by Alejandro Vaquero].}
\end{figure}

\begin{figure}
\begin{center}
\includegraphics[width=0.6\linewidth]{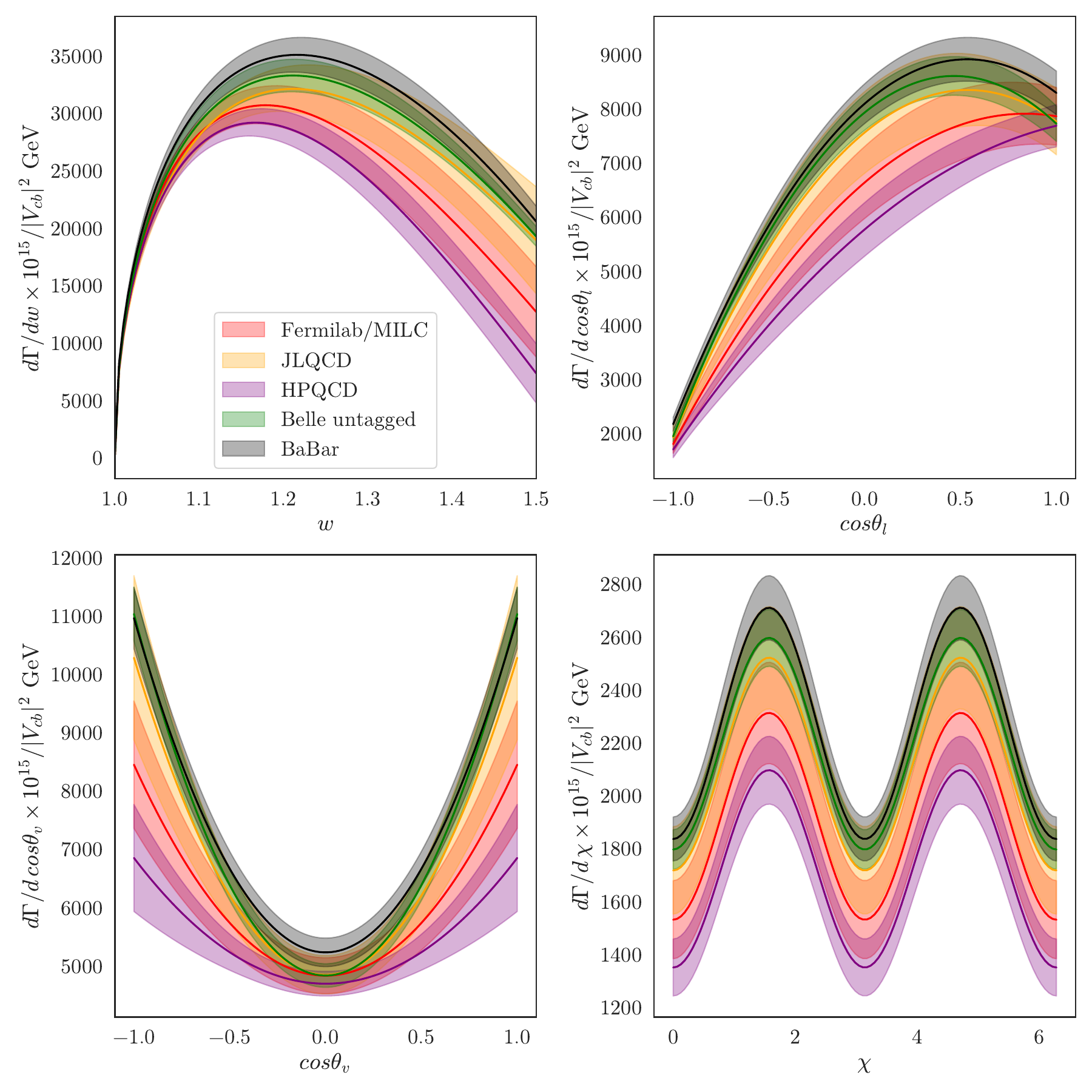}
\end{center}
\caption{\label{fig:BDstardecayrates}Comparison of lattice predictions \cite{FermilabLattice:2021cdg,Harrison:2023dzh,Aoki:2023qpa} and experimental results \cite{BaBar:2019vpl,Belle:2018ezy} for the $B\to D^*\ell\nu$ decay rate differential with respect to $w=v\cdot v^\prime$ (top left plot) and with respect to the decay angles (other three plots). The black and green curves are from BGL fits to the experimental data [plots by Alejandro Vaquero].}
\end{figure}

\begin{figure}
\begin{center}
\includegraphics[width=0.6\linewidth]{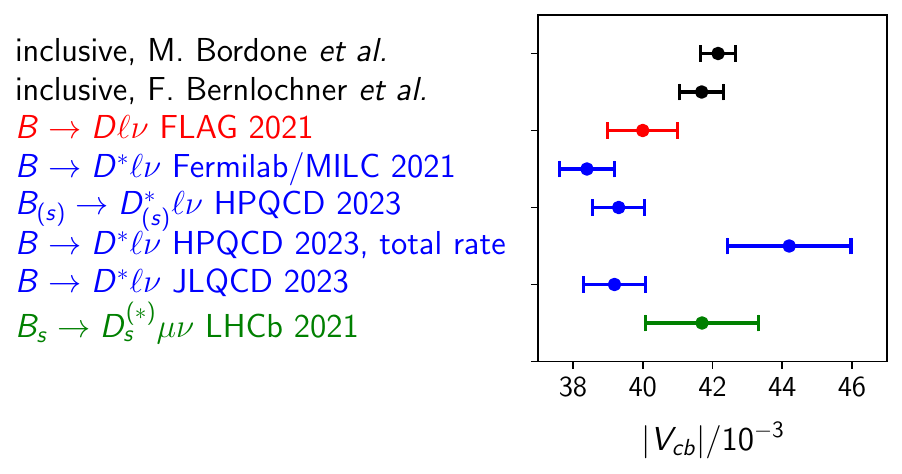} 
\end{center}
\caption{\label{fig:Vcb}Comparison of selected results for $|V_{cb}|$ as of July 2023: Inclusive, M.~Bordone {\it et al.} \cite{Bordone:2021oof}; Inclusive, F.~Bernlochner {\it et al.} \cite{Bernlochner:2022ucr} (the first extraction using $q^2$ moments); $B\to D\ell\nu$, FLAG \cite{FlavourLatticeAveragingGroupFLAG:2021npn} using form factors from  Fermilab/MILC \cite{MILC:2015uhg} and HPQCD \cite{Na:2015kha} and experimental data from Refs.~\cite{BaBar:2009zxk,Belle:2015pkj}; $B\to D^*\ell\nu$ using form factors from Fermilab/MILC \cite{FermilabLattice:2021cdg} and experimental data from Refs.~\cite{BaBar:2019vpl,Belle:2018ezy}, $B_{(s)}\to D_{(s)}^*\ell\nu$ using form factors from HPQCD \cite{Harrison:2023dzh} and experimental data from Refs.~\cite{Belle:2018ezy,LHCb:2020cyw};  $B\to D^*\ell\nu$ using form factors from JLQCD \cite{Aoki:2023qpa} and experimental data from Ref.~\cite{Belle:2018ezy}; $B_s\to D_s^{(*)}\mu\nu$, LHCb \cite{LHCb:2020cyw,LHCb:2021qbv} using form factors from HPQCD \cite{McLean:2019sds,McLean:2019qcx}.
}
\end{figure}

In 2023, two new lattice calculations of the $B\to D^*$ form factors were published, by the HPQCD \cite{Harrison:2023dzh} and JLQCD \cite{Aoki:2023qpa} collaborations. Like the 2021 Fermilab/MILC calculation \cite{FermilabLattice:2021cdg}, these works provide the full kinematic dependence of the vector and axial-vector form factors (past calculations for $B\to D^*$ were performed only at the zero-recoil point, where $q^2=q^2_{\rm max}$ and $w=v\cdot v^\prime=1$, meaning that experimental data had to be extrapolated to that point in order to extract $|V_{cb}|$). Table \ref{tab:BDstar} shows a comparison of the parameters of the three calculations. Notably, the HPQCD and JLQCD calculations use the same ``fully relativistic'' lattice action for the $b$ quark as used for the light, strange, and charm quarks, which simplifies the renormalization but requires extrapolations from unphysically light $b$-quark masses up to the physical value.

Comparisons of the $B\to D^*\ell\nu$ decay observables predicted using the lattice calculations in the Standard Model to experimental measurements \cite{BaBar:2019vpl,Belle:2018ezy} are shown in Figs.~\ref{fig:BDstarF} and \ref{fig:BDstardecayrates}. There is good agreement between the predictions using the JLQCD form factors and the measurements. The Fermilab/MILC and HPQCD calculations predict a somewhat steeper slope with respect to $w$ for the combination of form factors that appears in the differential decay rate. In addition, there are deviations between the SM predictions and measurements of the angular distributions, most significantly for HPQCD. Improved lattice calculations and new measurements by Belle II should help clarify the situation in the future.

A summary of results for $|V_{cb}|$ is shown in Fig.~\ref{fig:Vcb}. The values from $B\to D^*\ell\nu$ using combined fits to the lattice and experimental data performed by the three different lattice collaborations are consistent with each other and still significantly below the inclusive values \cite{Bordone:2021oof,Bernlochner:2022ucr}. Resolving this long-standing exclusive-inclusive discrepancy is of critical importance for flavor physics, since the uncertainty of the Wolfenstein parameter $A$ is the dominant source of uncertainty in Standard-Model predictions for many flavor-changing neutral-current processes. As is also shown in Fig.~\ref{fig:Vcb}, extracting $|V_{cb}|$ from the total rate measured in experiment and predicted by HPQCD's form factors instead of the combined fit to the differential distribution gives a much higher value of $|V_{cb}|$ (albeit with larger uncertainties); this is due to the discrepancy in the predicted slope.

At this conference, Andrew Lytle presented preliminary results for the $B_s\to D_s$ form factors using the fully relativistic all-HISQ approach from the Fermilab/MILC collaboration \cite{Lytle}, and Benjamin Choi and Seungyeob Jwa from the LANL-SWME collaboration reported on their progress toward a calculation of the $B\to D^*$ form factors using the Oktay-Kronfeld action for the heavy quarks on the MILC HISQ ensembles \cite{Bhattacharya:2023llf,Bhattacharya:2023ndo}. Aurora Melis presented a lattice determination of the susceptibilities that enter in dispersive bounds on $b\to c$ form factors \cite{Melis:2024wpb}. Alessandro Barone showed a comparison of the Chebyshev and Backus-Gilbert reconstruction methods applied to inclusive $B_s \to X_c \ell\nu$ decays, demonstrating that these methods yield consistent results \cite{Barone:2023iat}. Ryan Kellerman reported on a study of finite-volume effects in the Chebyshev approach \cite{Kellermann:2023yec}.

\FloatBarrier
\subsection{Determination of $\bss{\bar{\rho}}$ and $\bss{\bar{\eta}}$}

\begin{figure}[!h]
\begin{center}
 \includegraphics[width=0.4\linewidth]{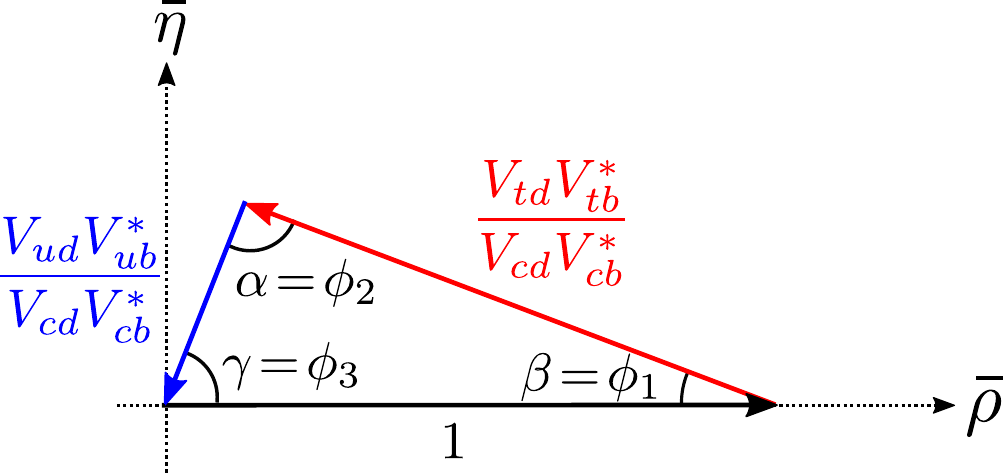}
\end{center}
\caption{\label{fig:triangle}The triangle, in the complex plane, representing the orthogonality of the first and third columns of the CKM matrix. }
\end{figure}

\noindent The remaining two Wolfenstein parameters are $\rho$ and $\eta$, or, to ensure exact unitarity, $\bar{\rho}$ and $\bar{\eta}$ \cite{PDGCKM}:
\begin{equation}
V^*_{ub}=A\lambda^3 (\rho + i \eta) = \frac{\sqrt{1-A^2\lambda^4}}{\sqrt{1-\lambda^2}[1-A^2\lambda^4(\bar{\rho}+i\bar{\eta})]} A\lambda^3(\bar{\rho}+i\bar{\eta}).
\end{equation}
Also note that $\bar{\rho}+i\bar{\eta}=-\displaystyle\frac{V_{ud}V_{ub}^*}{V_{cd}V_{cb}^*}$, and the orthogonality of the first and third columns of the CKM matrix, $V_{ud}V_{ub}^*+V_{cd}V_{cb}^*+V_{td}V_{tb}^*=0$, can be represented as a triangle in the complex plane with apex $\bar{\rho}+i\bar{\eta}$ as shown in Fig.~\ref{fig:triangle}.

\subsubsection{Determination of $\bss{|V_{ub}|}$}
\label{sec:Vub}

\noindent The magnitude {\large $|V_{ub}| = A \lambda^3 \sqrt{\rho^2+\eta^2}$ } can be determined from $b$-hadron semileptonic decays. The most important processes currently used to determine $|V_{ub}|$ are
\begin{itemize}\setlength{\itemsep}{0ex}
 \item Inclusive $B\to X_u \ell\nu$ ($\ell=e,\mu$; BaBar, Belle, Belle II, and older experiments)
 \item Exclusive $B\to \pi\ell\nu$ ($\ell=e,\mu$; BaBar, Belle, Belle II, and older experiments)
 \item Exclusive $B\to \rho \ell\nu$ and $B\to \omega \ell\nu$ ($\ell=e,\mu$; BaBar, Belle, Belle II, and older experiments, still using light-cone sum rules)
 \item Exclusive $B_s\to K\mu\nu$ (LCHb)
 \item Exclusive $\Lambda_b \to p\mu\bar{\nu}$ (LCHb)
 \item Exclusive $B\to \tau\nu$ (BaBar, Belle, Belle II, and older experiments)
\end{itemize}
The inclusive determination of $|V_{ub}|$ is more difficult compared to $|V_{cb}|$  due to the large $b\to c\ell\bar{\nu}$ background. Cutting away this contribution with a requirement on the lepton energy leaves only the endpoint region with $2 E_\ell/m_b\sim 1$, where the local HQE breaks down. In this region, one needs to use a light-cone OPE, such that the HQE parameters are replaced by nonlocal matrix elements, the so-called shape functions \cite{Gambino:2020jvv}. The exclusive determinations using $B\to \pi\ell\nu$, $B_s\to K\mu\nu$, $\Lambda_b \to p\mu\bar{\nu}$, $B\to \tau\nu$ use form factors and the $B$ decay constant from lattice QCD. Calculating the $B\to\rho$ and $B\to\omega$ form factors in lattice QCD in a rigorous way is more complicated because the $\rho$ and $\omega$ are resonances that are unstable under the strong interactions. As a two-body resonance, the $\rho$ is more tractable than the $\omega$ and work is already underway to calculate the $B\to\rho(\to \pi\pi)\ell\nu$ form factors in lattice QCD using the Lellouch-L\"uscher method \cite{Lellouch:2000pv,Briceno:2014uqa}, as shown in Luka Leskovec's plenary talk \cite{Leskovec:2024pzb}.

\begin{figure}
\begin{center}
 \includegraphics[width=0.6\linewidth]{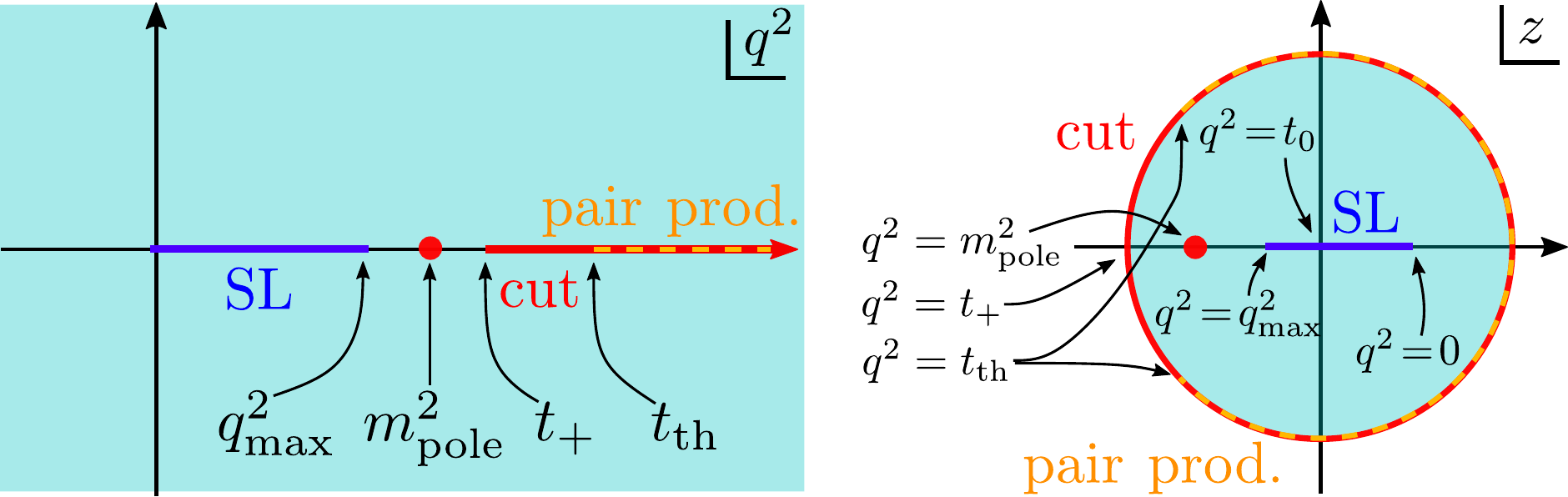}
\end{center}
\caption{\label{fig:zexpansion}The mapping of the complex $q^2$ plane to the unit disk in the complex $z$ plane through Eq.~(\protect\ref{eq:z}); see Ref.~\cite{Blake:2022vfl} for explanations. Note that Refs.~\cite{Flynn:2023qmi,Flynn:2023nhi} denote $t_+$ by $t^*$ and $t_{\rm th}$ by $t_+$.  }
\end{figure}

In 2023, there have been updates on the $B\to \pi$ and $B_s\to K$ form factors from lattice QCD. For $B\to \pi$, FLAG updated the average of lattice results by including the 2022 JLQCD results \cite{Colquhoun:2022atw} in their fit. The JLQCD calculation is the first to use the fully relativistic approach (with M\"obius domain-wall fermions in this case) and was already reviewed at Lattice 2022 \cite{Kaneko:2023kxx}. For $B_s\to K$, the RBC/UKQCD collaboration published an improved calculation \cite{Flynn:2023nhi,Hill} that supersedes their 2015 results \cite{Flynn:2015mha}, and I will discuss a preliminary update of the FLAG average below. Some of the plots in the following will show the form factors as a function of the variable $z$, which is defined as
\begin{equation}
 z(q^2)  = \frac{\sqrt{t_+ - q^2} -\sqrt{t_+ - t_0}}{\sqrt{t_+ - q^2} +\sqrt{t_+ - t_0}}\,, \label{eq:z}
\end{equation}
where $t_+$ and $t_0$ are constants. The properties of this change of variable are illustrated in Fig.~\ref{fig:zexpansion}. 
Furthermore, some of the plots show $B(q^2)f(q^2)$ instead of $f(q^2)$, where $B(q^2)=(1-m_{\rm pole}^2/q^2)$.

\begin{figure}

\vspace{4ex}

\footnotesize

\begin{center}
\begin{tabular}{|l|l|}
\hline
HPQCD 06 & 2+1 Asqtad, NRQCD $b$ (not included in fit) \\
FNAL/MILC 15 & 2+1 Asqtad, Fermilab $b$ \\
RBC/UKQCD 15 & 2+1 DWF, RHQ $b$ \\
JLQCD 22 & 2+1 DWF, DWF $b$ \\
\hline
\end{tabular}
\end{center}

\small

\begin{minipage}{0.49\linewidth}
\hspace{4.5ex} Old

\includegraphics[width=\linewidth]{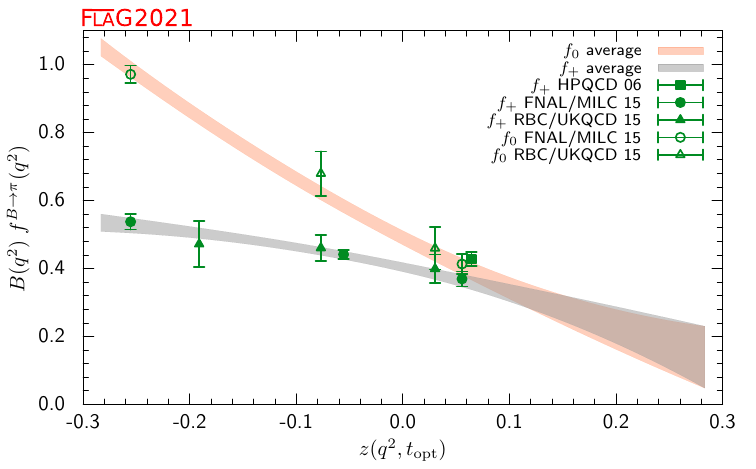}

\hspace{4.5ex} $\chi^2/{\rm dof}=0.82$

\phantom{ \hspace{5ex} All uncertainties rescaled by $\sqrt{\chi^2/{\rm dof}}$ }
\end{minipage}
\hfill
\begin{minipage}{0.49\linewidth}
\hspace{5ex}  New, including JLQCD 22

 \includegraphics[width=\linewidth]{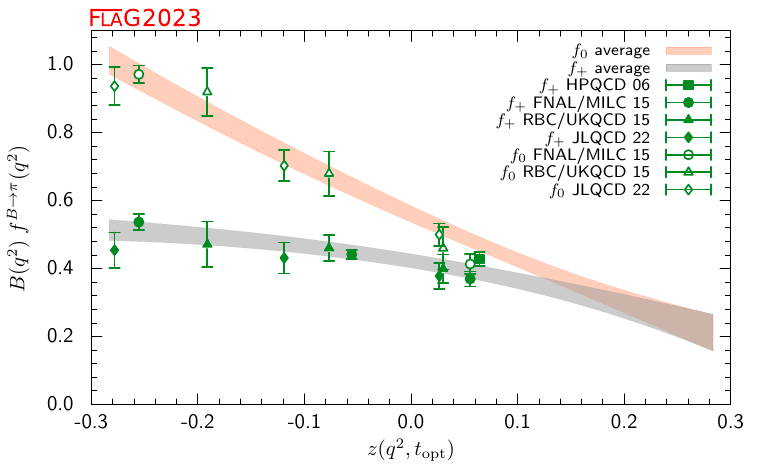}

 \hspace{5ex} \red{$\chi^2/{\rm dof}=3.63$}
 
 \hspace{5ex} All uncertainties rescaled by $\sqrt{\chi^2/{\rm dof}}$

\end{minipage}

\caption{\label{fig:B2pi}The effect of including the 2022 JLQCD results \cite{Colquhoun:2022atw} in the FLAG average of $B\to\pi$ form factors \cite{FlavourLatticeAveragingGroupFLAG:2021npn,FLAGweb}.}
\end{figure}

The effect of including the 2022 JLQCD results \cite{Colquhoun:2022atw} in the FLAG average of $B\to\pi$ form factors is shown in Fig.~\ref{fig:B2pi} \cite{FlavourLatticeAveragingGroupFLAG:2021npn,FLAGweb}. The updated BCL fit has a large $\chi^2/{\rm d.o.f.}$ caused by slight tensions between the results from the different collaborations; in particular in the slopes of $f_0$, which are very constrained due to strong correlations between the data points. The uncertainties of the fit parameters have therefore been rescaled by $\sqrt{\chi^2/{\rm d.o.f.}}$. Overall, the uncertainty of the $z$-expansion parameters $a_0^{0,+}$ has increased, while the uncertainty of $a_1^0$, $a_{1,2}^+$ has decreased. The result for $|V_{ub}|$ extracted from a combined fit of the lattice results and the BaBar \cite{BaBar:2010efp,BaBar:2012thb} and Belle \cite{Belle:2010hep,Belle:2013hlo} data has changed from $3.74(17)\times10^{-3}$ to $3.64(16)\times10^{-3}$ \cite{FlavourLatticeAveragingGroupFLAG:2021npn,FLAGweb}.

\begin{figure}
\begin{center}
\includegraphics[width=0.5\linewidth]{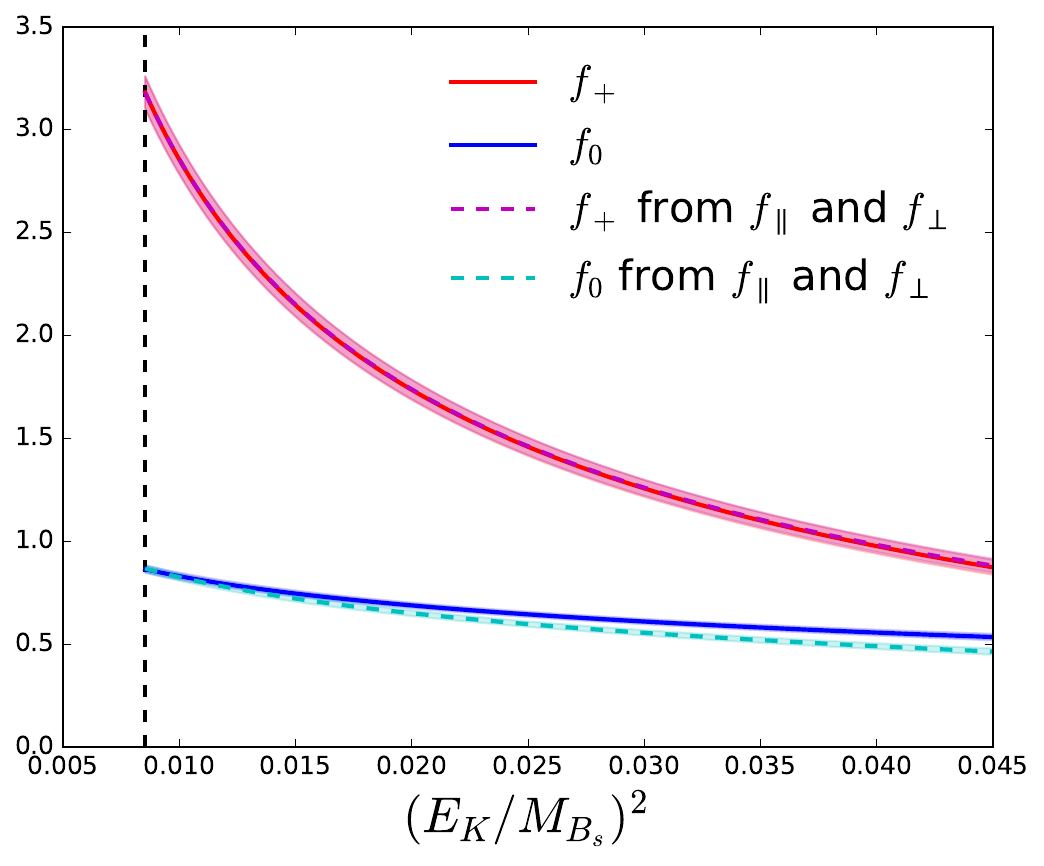} 
\end{center}
\caption{\label{fig:basisdepBsK}Results for the $B_s\to K$ form factors from chiral/continuum extrapolations performed directly for $f_+$ and $f_0$, compared to results obtained by performing the extrapolations for $f_\perp$ and $f_\parallel$ and then converting to $f_+$ and $f_0$ \cite{Flynn:2023nhi}.  }
\end{figure}

Coming to the $B_s \to K$ form factors, the new 2023 calculation by RBC/UKQCD \cite{Flynn:2023nhi,Hill} uses $N_f=2+1$ domain-wall fermions, RHQ (a.k.a. Columbia) anisotropic-clover $b$ quarks, and ``mostly nonperturbative'' renormalization like the 2015 calculation \cite{Flynn:2015mha}. The new calculation includes one additional ensemble with a third, finer lattice spacing that also has a lower pion mass than the other ensembles, uses updated scale setting and updated tuning of $m_s$ and of the RHQ parameters, and uses a different form-factor basis for the chiral-continuum extrapolations. These extrapolations use the fit functions \cite{Flynn:2023nhi}
\begin{align}
 \nonumber &f_X^{B_s\to K}(M_{\pi}, E_K, a^2) \\
  &=
  \frac{\Lambda}{E_K+\Delta_X}
  \left[c_{X,0}\bigg(1+
    \frac{\delta f(M^s_\pi)-\delta f(M^p_\pi)}{(4 \pi f_\pi)^2}\bigg) + c_{X,1} \frac{\Delta M_\pi^2}{\Lambda^2}  
  + c_{X,2}\frac{E_K}{\Lambda}
  + c_{X,3}\frac{E_K^2}{\Lambda^2} 
  + c_{X,4}(a\Lambda)^2 \right], \label{eq:pole}
\end{align}
where the first factor describes the pole from a $\bar{b}u$ bound state coupling to the weak current. The quantum numbers of this bound state are $J^P=1^-$ for the form factor $f_+$ and $J^P=0^+$ for the form factor $f_0$, so that $\Delta_+ = -42.1\:{\rm MeV}$ and $\Delta_0 = 263\:{\rm MeV}$ \cite{Flynn:2023nhi}. In the new RBC/UKQCD calculation, as well as in the 2014 HPQCD calculation \cite{Bouchard:2014ypa}, the chiral/continuum extrapolations were performed in this basis. In contrast, in the 2015 RBC/UKQCD calculation, as well as in the 2019 Fermilab/MILC calculation \cite{FermilabLattice:2019ikx}, the extrapolations were performed for $f_\perp$ and $f_\parallel$, each of which are linear combinations of $f_+$ and $f_0$, while still using the pole factor as in Eq.~(\ref{eq:pole}) with $\Delta_\perp=\Delta_+$ and $\Delta_\parallel=\Delta_0$; the justification for this choice was that $f_\perp$ is dominated by $f_+$ and $f_\parallel$ by $f_0$. The authors of the new RBC/UKQCD paper \cite{Flynn:2023nhi} compared both prescriptions and found a significant dependence on this basis choice in the final results for $f_0$, as shown here in Fig.~\ref{fig:basisdepBsK}.

\begin{figure}

\vspace{2ex}

\small

\begin{minipage}{0.49\linewidth}
\hspace{4.5ex} Old

\includegraphics[width=\linewidth]{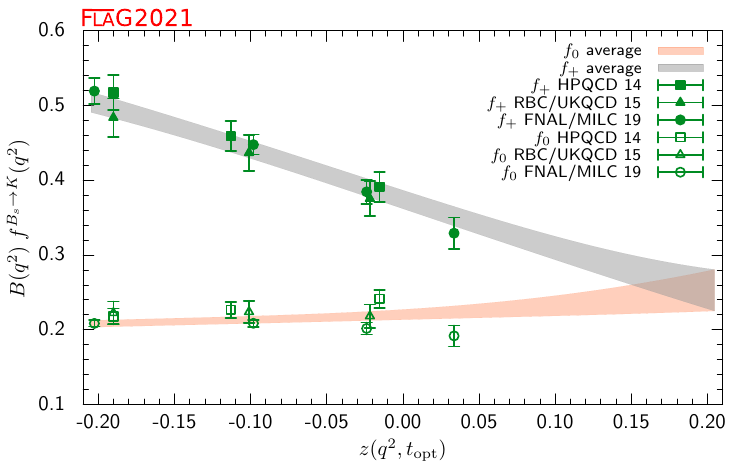}

\hspace{4.5ex} $\chi^2/{\rm dof}=1.54$

\hspace{4.5ex} All uncertainties rescaled by $\sqrt{\chi^2/{\rm dof}}$

\hspace{4.5ex}  $$\frac{\Gamma(B_s \to K \mu \nu)}{|V_{ub}|^2} = 6.28(0.67)\:\:{\rm ps}^{-1}$$

\end{minipage}
\hfill
\begin{minipage}{0.49\linewidth}

\hspace{5ex} New

\includegraphics[width=\linewidth]{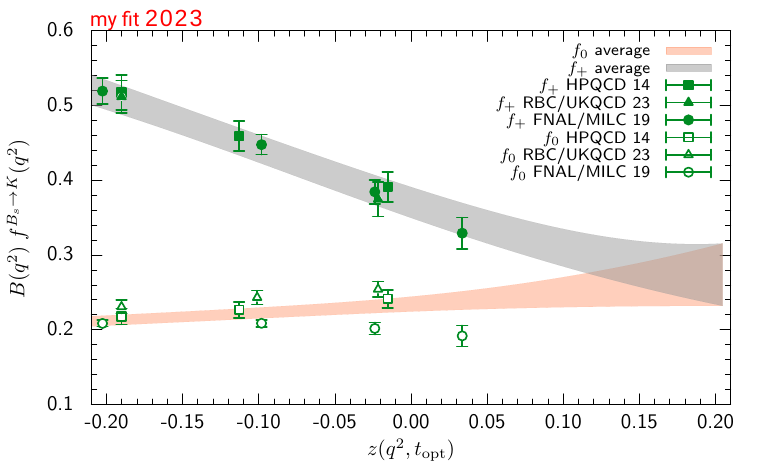}

\hspace{5ex} \red{$\chi^2/{\rm dof}=3.82$}

\hspace{5ex} All uncertainties rescaled by $\sqrt{\chi^2/{\rm dof}}$

\hspace{5ex}  $$\frac{\Gamma(B_s \to K \mu \nu)}{|V_{ub}|^2} = 6.5(1.1)\:\:{\rm ps}^{-1}$$

\end{minipage}

\caption{\label{fig:Bs2K}The effect of replacing the 2015 RBC/UKQCD results \cite{Flynn:2015mha} by the 2023 RBC/UKQCD results \cite{Flynn:2023nhi} in the FLAG average \cite{FlavourLatticeAveragingGroupFLAG:2021npn} of $B_s\to K$ form factors.}
\end{figure}

Figure \ref{fig:Bs2K} shows the effect of replacing the 2015 RBC/UKQCD results by the 2023 RBC/UKQCD results in the FLAG average of $B_s\to K$ form factors. The upward shift in the RBC/UKQCD results for $f_0$ leads to an increased tension between the lattice results from the different collaborations for this form factor. The increase in the scale factor $\sqrt{\chi^2/{\rm d.o.f.}}$ leads to an overall increase in the uncertainty of $\frac{\Gamma(B_s \to K \mu \nu)}{|V_{ub}|^2}$ calculated from the FLAG fit. The discrepancies in $f_0$ indicate that the uncertainties have been underestimated by at least some of the collaborations. It is possible that other sources of error besides the problem with extrapolating $f_\perp$ and $f_\parallel$ discussed above (which applies to the Fermilab/MILC results only), such as insufficiently controlled excited-state contamination, also contribute to the tension. Excited-state contamination can affect multiple steps of the analysis; it appears to have caused order-10\% biases in the $B_s$ kinetic-mass determinations in the tuning of the RHQ parameters by RBC/UKQCD, as discussed in Ref.~\cite{Meinel:2023wyg}.

\begin{figure}
\begin{center}
\includegraphics[width=0.6\linewidth]{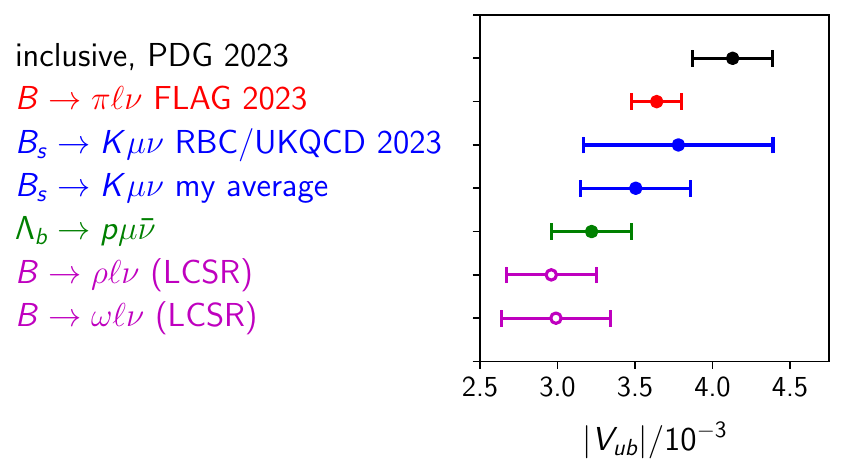} 
\end{center}

\caption{\label{fig:Vub}Comparison of selected results for $|V_{ub}|$ as of July 2023: from inclusive semileptonic $B$ decays \cite{PDGVubVcb}, from $B\to\pi\ell\nu$ \cite{FLAGweb} using form factors of Refs.~\cite{Flynn:2015mha,FermilabLattice:2015mwy,Colquhoun:2022atw} and experimental data from Refs.~\cite{BaBar:2010efp,BaBar:2012thb,Belle:2010hep,Belle:2013hlo}, from $B_s\to K\mu\nu$ using the form factors of Ref.~\cite{Flynn:2023nhi}, from $B_s\to K\mu\nu$ using my update of the FLAG average of the $B_s\to K$ form factors, from $\Lambda_b \to p \mu\bar{\nu}$ \cite{LHCb:2015eia} using the form factors of Ref.~\cite{Detmold:2015aaa}, and from $B\to\rho\ell\nu$, $B\to\omega\ell\nu$ \cite{Bernlochner:2021rel} using form factors from light-cone sum rules \cite{Bharucha:2015bzk}. Here, the determinations from $B_s\to K\mu\nu$ use the ratios $\mathcal{B}(B_s \to K \mu \nu)/\mathcal{B}(B_s \to D_s \mu \nu)$ and $\mathcal{B}(B_s \to D_s \mu \nu)/\mathcal{B}(B \to D \mu \nu)$ from LHCb \cite{LHCb:2020ist,LHCb:2020cyw} combined with $\mathcal{B}(B \to D \mu \nu)$ from PDG \cite{ParticleDataGroup:2022pth}, and the determination from $\Lambda_b\to p\mu\bar{\nu}$ uses the ratio $\mathcal{B}(\Lambda_b \to p\mu\bar{\nu})/\mathcal{B}(\Lambda_b \to \Lambda_c\mu\bar{\nu})$ from LHCb \cite{LHCb:2015eia} and  $|V_{cb}|=40.8(1.4)\times10^{-3}$ from PDG \cite{PDGVubVcb}.}
\end{figure}

A summary of $|V_{ub}|$ results is shown in Fig.~\ref{fig:Vub}. Note that the result from $B_s\to K\mu\nu$, RBC/UKQCD 2023 is based on secondary extrapolations of the form factors to the full $q^2$ range using a novel approach with unitarity bounds \cite{Flynn:2023qmi,Flynn:2023eok}, taking into account that the dispersive integral ranges only over an arc of the unit circle (in the $z$ plane) instead of the full circle for this process \cite{Blake:2022vfl}. However, these secondary extrapolations were not used in generating the inputs \cite{Flynn:2023nhi} to the FLAG fits in Fig.~\ref{fig:Bs2K}, and the FLAG fits also do not include unitarity bounds.

Several groups reported on their progress with new calculations of $b\to u$ semileptonic form factors at this conference. Protick Mohanta discussed JLQCD's study of the  $B_s\to K$ form factors using the M\"obius domain-wall action for all quarks \cite{Mohanta:2024hzi}. Andrew Lytle from the Fermilab/MILC collaboration presented preliminary results for the same form factors using an all-HISQ approach \cite{Lytle}, while Hwancheol Jeong reported on Fermilab/MILC's calculation of the $B\to\pi$ and $B_s\to K$ form factors using HISQ light quarks and Fermilab $b$ quarks \cite{Jeong}. I gave an update on a next-generation determination of the $\Lambda_b \to p$ form factors with domain-wall light quarks and RHQ $b$ quarks \cite{Meinel:2023wyg}. Luka Leskovec presented preliminary results for the $B\to\pi\pi$ ($I=1,L=1$) vector form factor at a heavier-than physical pion mass \cite{Leskovec:2024pzb}. Rainer Sommer discussed an approach for $b$-physics calculations on the lattice based on interpolations between relativistic and static computations and the step-scaling method \cite{Conigli:2023trw}, generalizing earlier work \cite{Guazzini:2007ja} to the case of semileptonic form factors and other observables. Alessandro Conigli presented preliminary numerical results for the $b$-quark mass and $B^{(*)}$-meson decay constants using this approach \cite{Conigli:2023rod}.

\subsubsection{Other constraints on the Wolfenstein parameters $\bss{\bar{\rho}}$ and $\bss{\bar{\eta}}$}

\begin{figure}
 
\begin{minipage}{0.6\linewidth}
 \centerline{\includegraphics[width=0.95\linewidth]{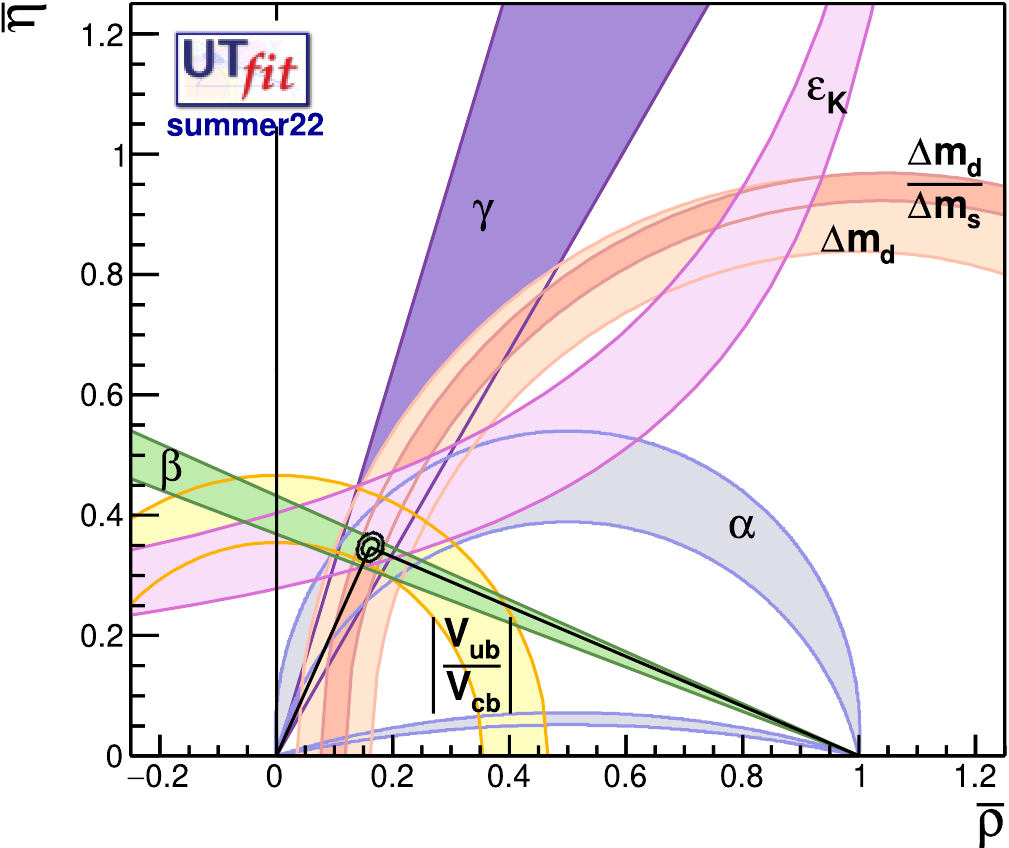}}
\end{minipage}
\begin{minipage}{0.39\linewidth}
 \scriptsize

 \begin{itemize}
  \item $\alpha$ from $CP$ violation in e.g.~$B^0(\overline{B^0})\to \pi\pi, \pi \rho, \rho\rho$
  \item $\beta$ from $CP$ violation in e.g.~$B^0(\overline{B^0})\to J/\psi K_S$
  \item $\gamma$ from $CP$ violation in e.g.~$B^-\to D^0(\overline{D^0})(\to f) K^-$
  \item $\Delta m_d$, $\displaystyle\frac{\Delta m_d}{\Delta m_s}$: $B^0/\overline{B^0}$, $B_s^0/\overline{B_s^0}$ mixing mass differences \magen{-- uses hadronic matrix elements from lattice QCD}
  \item $\epsilon_K$: indirect $CP$ violation in the neutral kaon system \magen{-- uses hadronic matrix elements from lattice QCD}
  \item $\epsilon^\prime_K$ (not shown): direct $CP$ violation in the neutral kaon system \magen{-- uses hadronic matrix elements from lattice QCD}
 \end{itemize}

\end{minipage}

\caption{\label{fig:UTfit}Constraints on the Wolfenstein parameters $\bar{\rho}$ and $\bar{\eta}$. Figure from UTfit \cite{UTfit:2022hsi} (see also Ref.~\cite{CKMfitter} for a similar analysis by the CKMfitter Group).}

\end{figure}

\noindent The determinations of $|V_{ub}|$ discussed in the previous section set the length of the left side of the unitarity triangle and constrain the apex to lie within the yellow ring in Fig.~\ref{fig:UTfit}. We will now discuss the other constraints shown in the figure.

The determinations of the angles $\alpha$, $\beta$, and $\gamma$ are based from CP-violation measurements in nonleptonic decays, where all necessary combinations of hadronic matrix elements can be determined by fitting experimental data.

\begin{figure}
\vspace{4ex}
\begin{center}
 \includegraphics[width=0.6\linewidth]{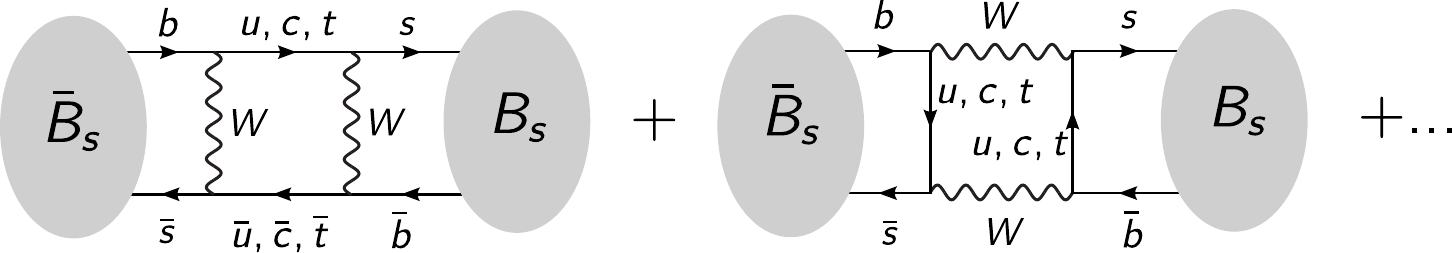}
\end{center}
\caption{\label{fig:Bmixing}Examples of diagrams contributing to $B_s^0$-$\overline{B_s^0}$ mixing in the Standard Model.}
\end{figure}

The length of the right side of the triangle is constrained by the $B^0_{(s)}$-$\overline{B^0_{(s)}}$ oscillation frequencies, which have been measured very precisely to be \cite{HFLAV:2022esi,HFLAVweb}
\begin{align}
 \nonumber \Delta m_d &= 0.5065(19)\:{\rm ps}^{-1}, \\
 \Delta m_s &= 17.765(6)\:{\rm ps}^{-1}.
\end{align}
Examples of diagrams contributing to $B_s^0$-$\overline{B_s^0}$ mixing in the Standard Model are shown in Fig.~\ref{fig:Bmixing}. The hadronic matrix elements currently taken from lattice QCD to relate the oscillation frequencies to the CKM parameters are
\begin{equation}
 \langle \overline{B^0_q} | O^{\Delta B=2}_q | B^0_q\rangle= \frac{8}{3}f_{B_q}^2 m_{B_q}^2 B_{B_q},\hspace{2ex} \text{where} \hspace{2ex} O^{\Delta B=2}_q=[\bar{b}\gamma_\mu(1-\gamma_5)q][\bar{b}\gamma^\mu(1-\gamma_5)q]. \label{eq:Bbag}
\end{equation}

The kaon CP violation parameters $\epsilon_K$ (indirect) and $\epsilon'_K$ (direct) are related to the ratios of $K\to\pi\pi$ amplitudes through \cite{PDGkaonCP}
\begin{align}
\frac{A(K^0_L \to \pi^+ \pi^-)}{A(K^0_S \to \pi^+ \pi^-)} \approx \epsilon_K + \epsilon'_K, \hspace{4ex} \frac{A(K^0_L \to \pi^0 \pi^0)}{A(K^0_S \to \pi^0 \pi^0)} \approx \epsilon_K - 2\epsilon'_K.
\end{align}
The measured values of these parameters are \cite{PDGkaonCP}
\begin{align}
 \nonumber \epsilon_K &= 2.228(11)\times 10^{-3}, \\
\mathrm{Re}(\epsilon^\prime_K/\epsilon_K) &= 1.66(23)\times 10^{-3}.
\end{align}
The hadronic matrix elements currently taken from lattice QCD for the Standard-Model calculation of $\epsilon_K$ and $\epsilon^\prime_K$ (and hence to constrain the CKM parameters using these observables) are
\begin{equation}
 \langle \pi\pi | O_i^{\Delta S=1} | K^0\rangle
\end{equation}
for seven different four-quark operators $O_i^{\Delta S=1}$ and
\begin{equation}
 \langle \overline{K^0} | O^{\Delta S=2} | K^0\rangle =\frac{8}{3}f_K^2 m_K^2 B_K, \hspace{2ex}\text{where}\hspace{2ex}O^{\Delta S=2}=[\bar{s}\gamma_\mu(1-\gamma_5)d][\bar{s}\gamma^\mu(1-\gamma_5)d]. \label{eq:Kbag}
\end{equation}
The kaon CP-violation parameters also receive contributions from nonlocal two-current matrix elements; the UTfit collaboration \cite{UTfit:2022hsi} currently takes these matrix elements from a chiral-perturbation-theory calculation \cite{Buras:2010pza}, 
but they can, in principle, also be calculated in lattice QCD \cite{Christ:2011fly,Bai:2014cva,Wang:2022lfq,Jackura:2022xml}.

It is worth noting that a major source of uncertainty in the Standard-Model predictions of $\epsilon_K$ and $\Delta m_d$, and hence in the uncertainties of the corresponding constraints on $\bar{\rho}$ and $\bar{\eta}$ (and constraints on new physics), is the uncertainty of the Wolfenstein parameter $A$, i.e., the uncertainty of $|V_{cb}|$ \cite{Jwa:2023efi,Charles:2020dfl}.

\begin{figure}
\includegraphics[width=0.45\linewidth]{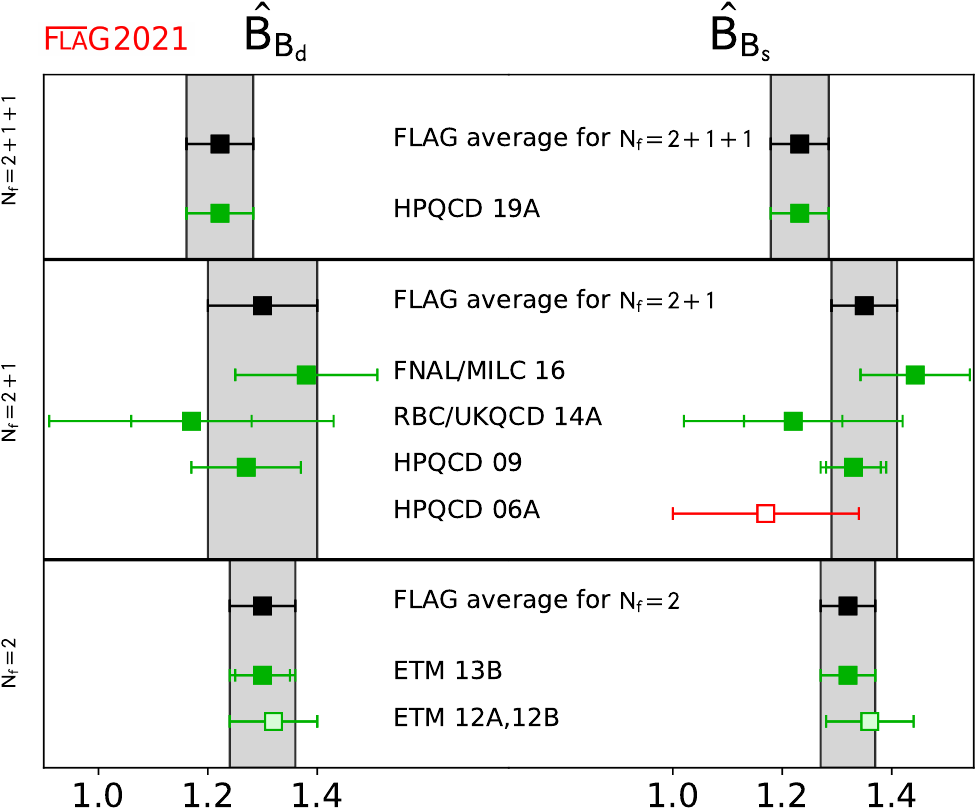} \hfill \includegraphics[width=0.45\linewidth]{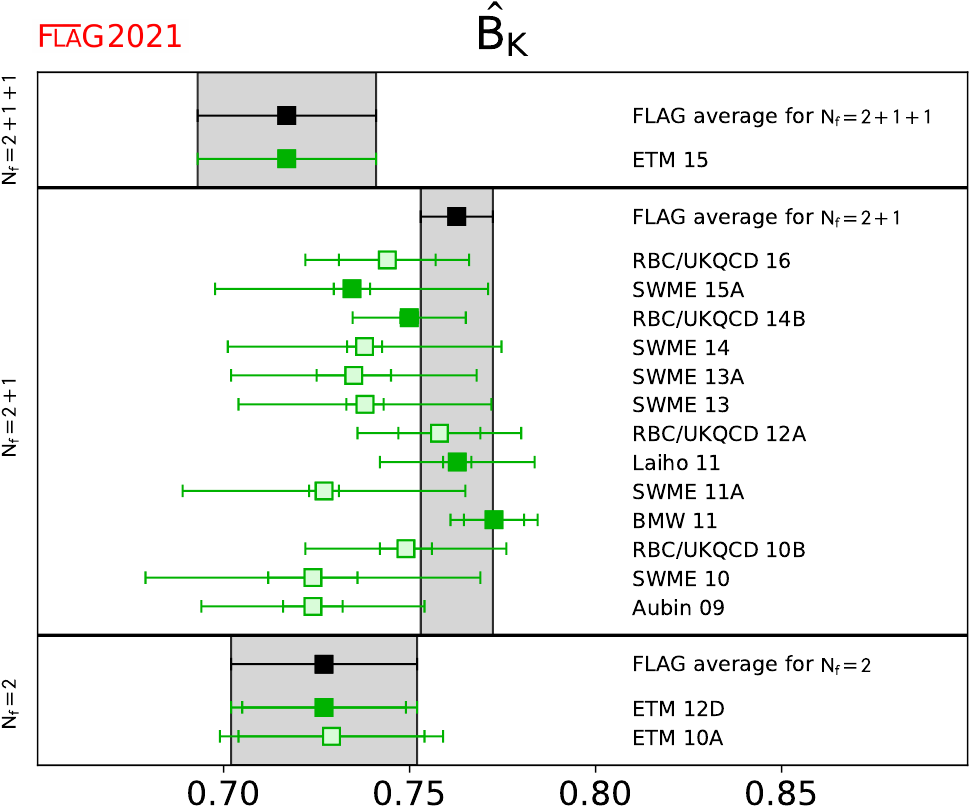}
\caption{\label{fig:bagparams}FLAG summary of lattice results for the $B$, $B_s$, and $K$ meson renormalization-group-independent bag parameters \cite{FlavourLatticeAveragingGroupFLAG:2021npn}.}
\end{figure}

A summary of lattice results for the renormalization-group-independent versions \cite{FlavourLatticeAveragingGroupFLAG:2021npn} of the bag parameters defined in Eqs.~(\ref{eq:Bbag}) and (\ref{eq:Kbag}) is shown in Fig.~\ref{fig:bagparams}. No new complete lattice results for the kaon bag parameter have been published since 2016. A major challenge and source of systematic uncertainty in lattice calculations of four-quark-operator matrix elements is the renormalization and matching, and it is good to see progress in this area at this conference, as discussed later. The $K\to(\pi\pi)_{I=2}$ and $K\to(\pi\pi)_{I=0}$ matrix elements were computed in Ref.s~\cite{Blum:2015ywa} and \cite{RBC:2020kdj}, respectively. The combination of the isospin-2 and isospin-0 amplitudes give the Standard-Model prediction $\mathrm{Re}(\epsilon^\prime_K/\epsilon_K)_{\rm SM}=  2.17(26)(62)(50) \times 10^{-3}$ \cite{RBC:2020kdj}. In these calculations, another big challenge is the precise determination of the $\pi\pi$ finite-volume spectrum and the corresponding Lellouch-L\"uscher factors. Reference \cite{RBC:2020kdj} used so-called $G$-parity boundary conditions to obtain physical kinematics.

At this conference, Justus Tobias Tsang presented an ongoing new calculation of the $B_s^0$-$\overline{B_s^0}$ mixing matrix elements using a domain-wall action for all quarks and RI/sMOM renormalization \cite{Tsang}. The calculation uses both RBC/UKQCD and JLQCD ensembles. Matthew Black discussed how the gradient flow and short-flow-time expansion can be used to renormalize four-quark operators, including both the $\Delta B=2$ operators relevant for neutral-meson mixing and the $\Delta B=0$ operators relevant for the theory of heavy-hadron lifetimes \cite{Black:2023vju}. Joshua Lin presented a position-space renormalization scheme for $\Delta B=2$ and $\Delta B=0$ four-quark operators in HQET and reported on progress with the matching from this scheme to $\overline{\text{MS}}$ at order $\alpha_s$ \cite{JoshLin}. Riccardo Marinelli presented a nonperturbative calculation of the renormalization-group running of the complete basis of $\Delta S=2$ four-quark operators in the chirally rotated Schrödinger functional in the region of energies between the $W$ mass and the switching scale of 4 GeV; this running is traditionally done perturbatively, and the new method can eliminate the associated uncertainty \cite{Marinelli}. Weonjong Lee gave an update on the Standard-Model prediction for $\epsilon_K$ using lattice inputs, where a significant tension with the experimental value is observed when using $|V_{cb}|$ from exclusive decays, and this tension disappears when using the inclusively determined $|V_{cb}|$ instead \cite{Jwa:2023efi}. Masaaki Tomii presented a new RBC/UKQCD calculation of the $K\to\pi\pi$ matrix elements with periodic instead of $G$-parity boundary conditions, using the variational method to extract the matrix elements for excited $\pi\pi$ finite-volume states to cover the energy region around the physical kinematic point \cite{Tomii,RBC:2023ynh}.

\subsection{Global fit of the Wolfenstein parameters}

\noindent To conclude this part, I quote the final results for all four Wolfenstein parameters
from a global fit to the observables discussed in the previous sections. The 2022 Standard-Model fit by UTfit gives \cite{UTfit:2022hsi}
\begin{eqnarray}
\nonumber \lambda  &=& 0.22519(83), \\
\nonumber A &=& 0.828 (11), \\
\nonumber \bar\rho &=& 0.161 (10), \\
 \bar\eta &=& 0.347 (10),
\end{eqnarray}
which corresponds to the CKM matrix elements

\begin{align}
\nonumber & \left(\! \begin{array}{lll} V_{ud}  & V_{us} & V_{ub} \\ V_{cd} & V_{cs} & V_{cb} \\ V_{td} & V_{ts} & V_{tb} \end{array}\!  \right) \\
&= \left(\!\! \begin{array}{lll}
  \wm0.97431(19)    &  \wm0.22517(81) & 0.003715(93)\, e^{-i (65.1(1.3))^{\circ}} \\
  -0.22503(83) \, e^{+i (0.0351(1))^{\circ}}   &  \wm0.97345(20)\, e^{-i (0.00187(5))^{\circ}} & 0.0420(5)  \\
  \wm0.00859 (11)\, e^{-i (22.4(7))^{\circ}} & -0.04128 (46)  \, e^{+i (1.05(3))^{\circ}} & 0.999111(20)
 \end{array} \right). \label{eq:CKMglobalfit}
\end{align}
See also Refs.~\cite{CKMfitter,Hocker:2001xe,Charles:2015gya} for a similar analysis by the CKMfitter Group.

\FloatBarrier
\section{Selected further processes in quark flavor physics}

\subsection{Tests of lepton-flavor universality in $\bss{b\to c\ell\bar{\nu}}$ decays}

\begin{figure}
\begin{center}
 \includegraphics[height=4cm]{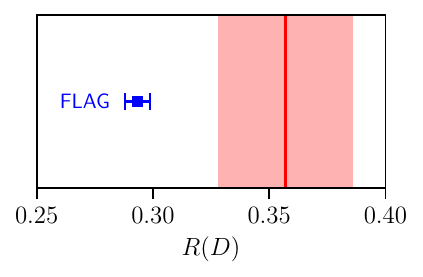} \hspace{1ex} \includegraphics[height=4cm]{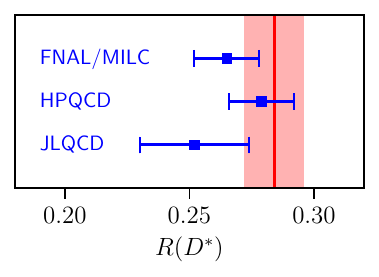}

 \hspace{1ex} \includegraphics[height=4cm]{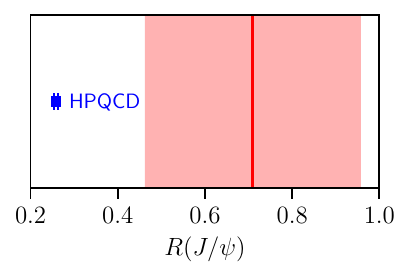} \hspace{0.5ex} \includegraphics[height=4cm]{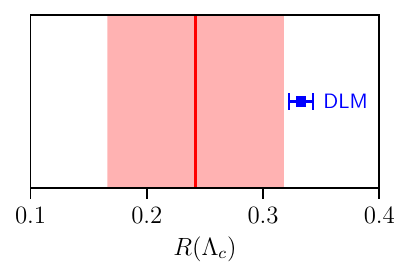}
\end{center}

\caption{\label{fig:RHc}Comparison of lattice-only SM predictions (blue points with error bars) and experimental measurements (red lines with error bands) of the ratios $R(H_c)$.
For $R(D)$, the SM prediction from FLAG \cite{FlavourLatticeAveragingGroupFLAG:2021npn} uses the average of form factors from Fermilab/MILC \cite{MILC:2015uhg} and HPQCD \cite{Na:2015kha}; the experimental value is the HFLAV average \cite{HFLAVRD} of measurement by BaBar \cite{BaBar:2012obs}, Belle \cite{Belle:2015qfa,Belle:2019rba}, and LHCb \cite{LHCb:2023zxo}. For $R(D^*)$, the SM predictions are from Fermilab/MILC \cite{FermilabLattice:2021cdg}, HPQCD \cite{Harrison:2023dzh}, and JLQCD \cite{Aoki:2023qpa}, and the experimental value is the HFLAV average \cite{HFLAVRD} of measurements by BaBar \cite{BaBar:2012obs}, Belle \cite{Belle:2015qfa,Belle:2016dyj,Belle:2019rba}, Belle II \cite{Belle-II:2024ami}, and LHCb \cite{LHCb:2023zxo,LHCb:2023uiv}. For $R(J/\psi)$, the SM prediction is from HPQCD \cite{Harrison:2020nrv} and the measurement from LHCb \cite{LHCb:2017vlu}. For $R(\Lambda_c)$, the SM prediction is from Detmold, Lehner, and Meinel \cite{Detmold:2015aaa} and the measurement is from LHCb \cite{LHCb:2022piu}.}
\end{figure}

\noindent In the Standard Model, the different generations of leptons couple to $W$ bosons with the same strength, and it is interesting to search for deviations from this lepton-flavor universality. Important observables for such tests are the ratios
\begin{equation}
R(H_c)=\frac{\Gamma(H_b\to H_c \tau \nu)}{\Gamma(H_b\to H_c \ell \nu)},
\end{equation}
where $H_b$ denotes a bottom hadron, $H_c$ a charm hadron, and $\ell$ an electron or muon. The SM predictions of $R(H_c)$ depend on the $H_b\to H_c$ form factors, some of which were already discussed in Sec.~\ref{sec:A}. A comparison of experimental measurements and SM predictions using form factors from lattice QCD is shown in Fig.~\ref{fig:RHc}. For $R(D^*)$, the pure SM predictions that do not rely on experimental information on the $B\to D^*\ell\nu$ decay distribution became available only relatively recently \cite{FermilabLattice:2021cdg,Harrison:2023dzh,Aoki:2023qpa}. The measurements of $R(H_c)$ with mesons show hints for excesses with respect to the SM predictions; these hints were first seen in 2012 \cite{BaBar:2012obs} and many possible new-physics explanations have been proposed. In contrast, the recent LHCb result for the baryon-decay ratio $R(\Lambda_c)$ \cite{LHCb:2022piu} has a central value below the SM prediction \cite{Detmold:2015aaa}. No common explanation of the measured values of all ratios within $2\sigma$ using heavy new physics is possible \cite{Fedele:2022iib}.

\subsection{Direct determinations of $\bss{|V_{cd}|}$ and $\bss{|V_{cs}|}$}

\noindent With $|V_{cd}|$ and $|V_{cs}|$ predicted precisely by the global fit (without inputs from charm decays) as shown in Eq.~(\ref{eq:CKMglobalfit}), it is interesting to check whether direct determinations are compatible with these values. Experimental data are more precise for semileptonic $D_{(s)}$ decays compared to leptonic $D_{(s)}$ decays. In the past, leptonic decays nevertheless gave the most precise $|V_{cs}|$ and $|V_{cd}|$ because lattice results for decay constants are more precise than for form factors. Recently, lattice results for $D_{(s)}$ semileptonic decays have reached high precision \cite{Chakraborty:2021qav,FermilabLattice:2022gku}, and semileptonic decays currently yield the most precise direct $|V_{cs}|$ and $|V_{cd}|$ determinations (but there are some tensions with older lattice results as discussed below).

\begin{figure}
 \includegraphics[width=0.49\linewidth]{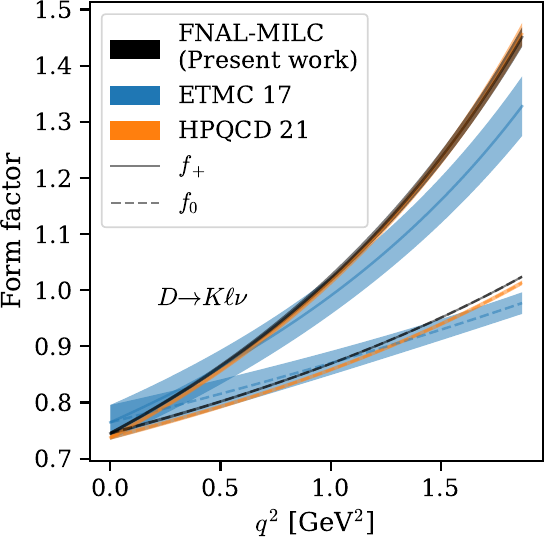} \hfill \includegraphics[width=0.49\linewidth]{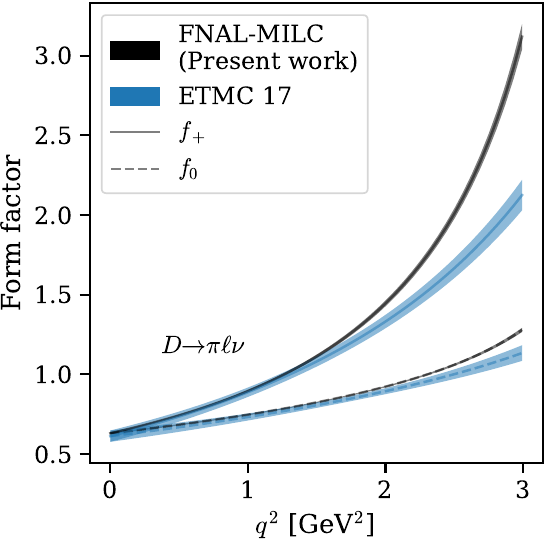}
 \caption{\label{fig:DSLcomp}Comparison of Fermilab/MILC's new results for the $D\to K$ and $D\to\pi$ form factors \cite{FermilabLattice:2022gku} to previous results from HPQCD \cite{Chakraborty:2021qav} and ETMC \cite{Lubicz:2017syv}. Figure from Ref.~\cite{FermilabLattice:2022gku}. }
\end{figure}

In the year leading up to the Lattice conference, the Fermilab/MILC collaboration published a new calculation of the $D\to\pi$ and $D_{(s)}\to K$ form factors using the $N_f=2+1+1$ all-HISQ approach \cite{FermilabLattice:2022gku}. A comparison of these results with the ones from HPQCD \cite{Chakraborty:2021qav} and ETMC \cite{Lubicz:2017syv} is shown in Fig.~\ref{fig:DSLcomp}. Good agreement is seen for the form factor $f_+(D\to K)$ that is most relevant for the $D\to K\ell\nu$ decay rate between Fermilab/MILC and HPQCD, while for $f_0(D\to K)$, there appears to be a small but statistically significant tension between the Fermilab/MILC and HPQCD results. Compared to the older ETMC results that are based on computations with the twisted-mass and Osterwalder-Seiler actions \cite{Lubicz:2017syv}, the Fermilab/MILC results for all form factors are higher in the high-$q^2$ region, most dramatically for $f_+(D\to\pi)$. Even though this kinematic region contributes less to the decay rate than the low-$q^2$ region, it would be good to understand the origin of this large discrepancy.

\begin{figure}
\begin{center}
 \includegraphics[width=0.8\linewidth]{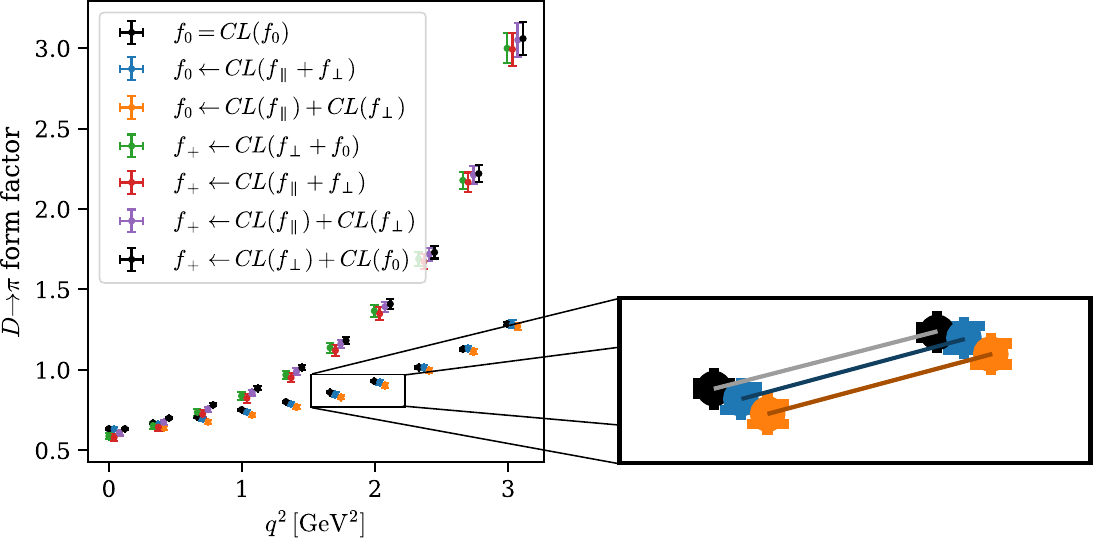}
\caption{\label{fig:Dpibasis}Dependence of the Fermilab/MILC $D\to\pi$ form-factor results on the choice of basis for the chiral-continuum extrapolations \cite{FermilabLattice:2022gku}. Figure modified from Ref.~\cite{FermilabLattice:2022gku}; I added the magnification box.}
 \end{center}
 \end{figure}

The Fermilab/MILC collaboration also investigated the dependence of the form-factor results on the choice of basis used for the chiral and continuum extrapolation (cf.~Sec.~\ref{sec:Vub}), as shown in Fig.~\ref{fig:Dpibasis}. Similarly to the findings for $B_s\to K$ by RBC/UKQCD \cite{Flynn:2023nhi}, doing the extrapolation directly for $f_0$ gives slightly higher values for this form factor compared to constructing $f_0$ from the extrapolated $f_\perp$ and $f_\parallel$ (I thank Andreas J\"uttner for pointing this out to me).

\begin{figure}
 \includegraphics[width=0.45\linewidth]{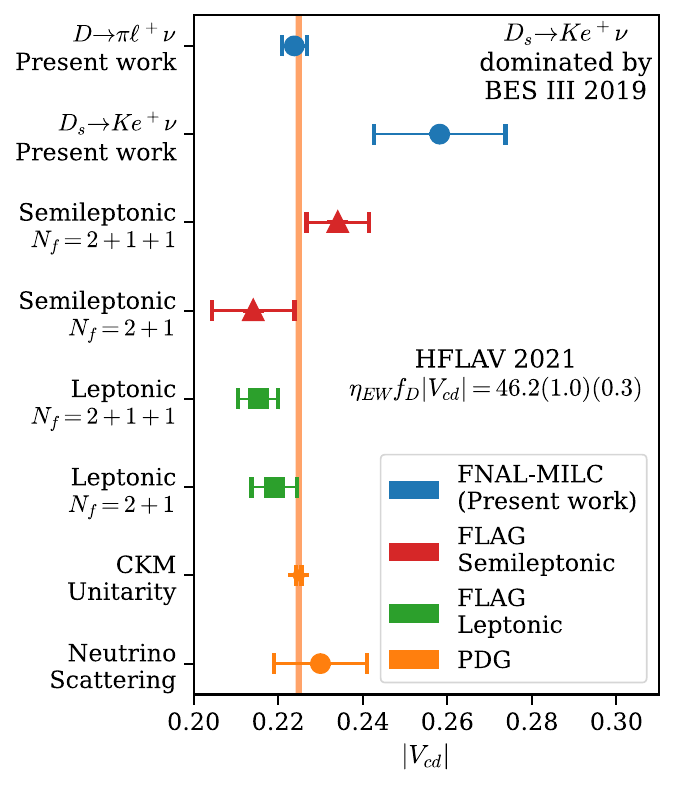} \hfill \includegraphics[width=0.45\linewidth]{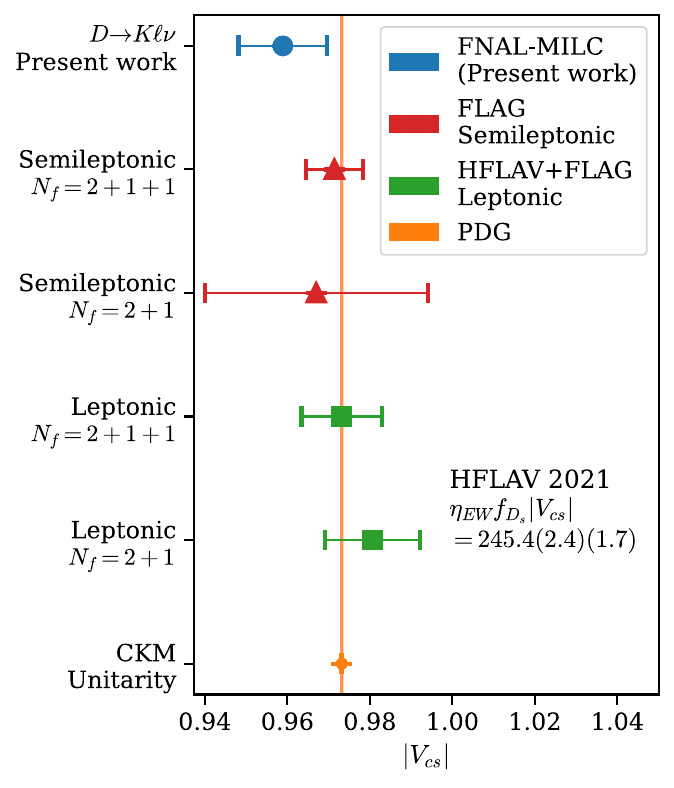}
 \caption{\label{fig:VcdVcs}Comparison of direct $|V_{cd}|$ and $|V_{cs}|$ determinations from meson decays and neutrino scattering \cite{FermilabLattice:2022gku,FlavourLatticeAveragingGroupFLAG:2021npn,ParticleDataGroup:2022pth} with the values predicted by the global CKM fit using unitarity \cite{UTfit:2022hsi}. The experimental data for the semileptonic decays are from BaBar \cite{BaBar:2007zgf,BaBar:2014xzf}, CLEO \cite{CLEO:2009svp}, and BES III \cite{BESIII:2015tql,BESIII:2017ylw,BESIII:2018nzb,BESIII:2018ccy,BESIII:2018xre}. Figure modified from Ref.~\cite{FermilabLattice:2022gku}; I removed the blue bands and added the orange unitarity bands. In addition to what is shown here, there are also $|V_{cs}|$ results from $\Lambda_c\to\Lambda\ell\nu$, most recently $|V_{cs}|=0.937(29)$ \cite{Meinel:2016dqj,BESIII:2023vfi}.}
\end{figure}

A comparison of the most recent direct $|V_{cd}|$ and $|V_{cs}|$ determinations with the values predicted by the global CKM fit using unitarity is shown in Fig.~\ref{fig:VcdVcs}. The results from the semileptonic decays are consistent with unitarity within 1-2$\sigma$.

\begin{figure}
 
\begin{minipage}{0.59\linewidth}
 \includegraphics[width=\linewidth]{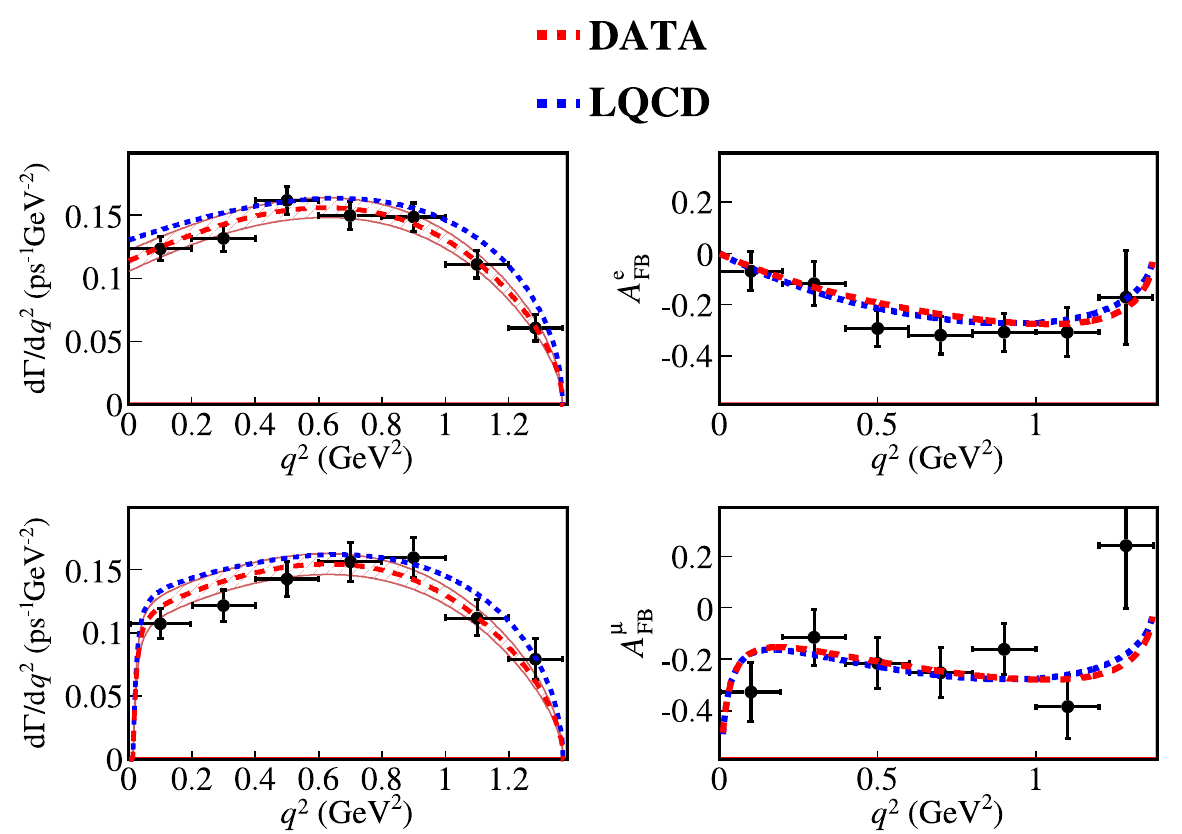}
\end{minipage}
\hfill
\begin{minipage}{0.35\linewidth}
 \includegraphics[width=\linewidth]{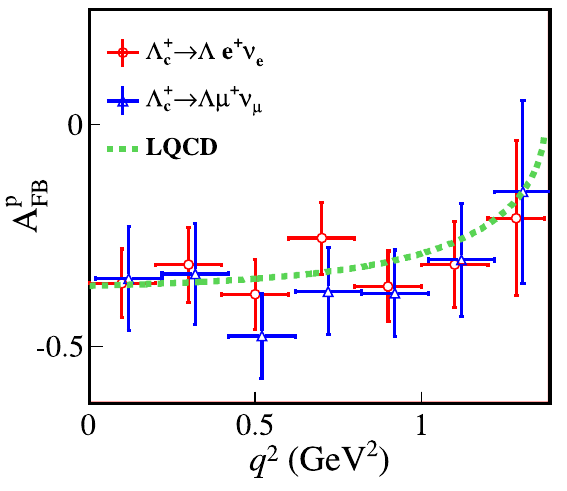}
\end{minipage}

\begin{center}

 \includegraphics[width=0.6\linewidth]{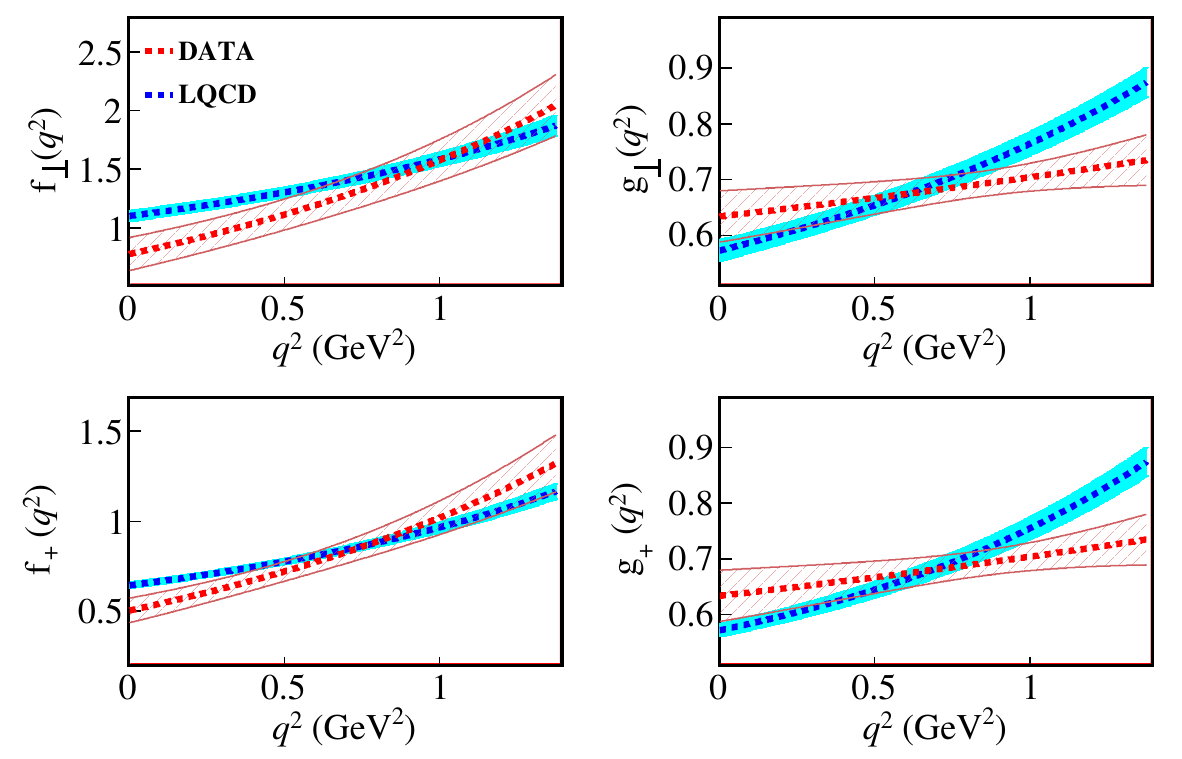}
\end{center}

\caption{\label{fig:LcL}Upper five panels: BESIII binned measurements and fits of the $\Lambda_c\to\Lambda e^+\nu$ and $\Lambda_c\to\Lambda \mu^+\nu$ differential decay rates, lepton-side forward-backward asymmetries, and hadron-side forward-backward asymmetries \cite{BESIII:2023vfi}, compared to Standard-Model predictions (only central values shown) using form factors from lattice QCD \cite{Meinel:2016dqj}. Lower four panels: the form-factor parametrization fitted to the BESIII $\Lambda_c\to\Lambda(\to p \pi)\ell^+\nu$ angular-distribution data \cite{BESIII:2023vfi} compared to the predictions from lattice QCD \cite{Meinel:2016dqj}. All plots from Ref.~\cite{BESIII:2023vfi}.}

\end{figure}

The $c\to s\ell^+\nu$ transition can also be observed in baryon decays. In 2023, the BESIII collaboration published a combined analysis of the $\Lambda_c\to\Lambda e^+\nu$ and $\Lambda_c\to\Lambda \mu^+\nu$ differential branching fractions and angular distributions \cite{BESIII:2023vfi}, updating their 2022 analysis of $\Lambda_c\to\Lambda e^+\nu$ only \cite{BESIII:2022ysa}. The results are compared to the SM predictions using lattice QCD \cite{Meinel:2016dqj} in Fig.~\ref{fig:LcL}. From the total rates, the result $|V_{cs}| =0.937\pm0.014_{\mathcal{B}}\pm0.024_{\rm LQCD}\pm0.007_{\tau_{\Lambda_c}}$ is obtained \cite{BESIII:2023vfi}, where the uncertainty from the lattice calculation of the form factors now dominates. In addition, BESIII fitted a simplified form-factor parametrization to the $\Lambda_c\to\Lambda(\to p \pi)\ell^+\nu$ angular distributions, as shown in the lower panels in Fig.~\ref{fig:LcL}. Compared to my lattice-QCD predictions \cite{Meinel:2016dqj}, the BESIII fit yields slightly steeper slopes for the vector form factors and less steep slopes for the axial-vector form factors (there is also a recent independent lattice calculation of the $\Lambda_c\to\Lambda$ form factors \cite{Bahtiyar:2021voz}, but it used only a single $N_f=2$ ensemble on a $16^3\times 32$ lattice with $a\approx 0.16$ fm, $m_\pi\approx 550$ MeV).

At this conference, Callum Farrell presented preliminary lattice-QCD results for another charm-baryon semileptonic decay, $\Xi_c\to\Xi\ell^+\nu$ \cite{Farrell:2023vnm}. The calculation points toward a branching fraction in the Standard Model that is substantially higher than the current experimental results, and also higher than predicted by a previous lattice calculation \cite{Zhang:2021oja}. Davide Giusti presented a lattice calculation of the structure-dependent form factors in the radiative leptonic decays $D_s\to \ell\nu\gamma$ \cite{Giusti:2023pot,DavideGiusti}. Ryan Kellermann reported on a study of finite-volume effects in a lattice determination of the inclusive $D_s\to X_s\ell\nu$ rate \cite{Kellermann:2023yec}. Antonin Portelli discussed the occurrence of finite-volume collinear divergences in the $\text{QED}_L$ treatment of electromagnetic corrections to $D_{(s)}$ leptonic decays \cite{Portelli}. Justus Kuhlmann \cite{Fritzsch:2024zux} presented an exploration of stabilized Wilson fermions \cite{Francis:2019muy} for charm physics.

\subsection{Rare $\bss{b\to s \ell^+\ell^-}$ decays}

\noindent Decays of $b$-hadrons involving the transition $b\to s \ell^+\ell^-$ are among the most important modes used to search for physics beyond the Standard Model \cite{Altmannshofer:2022hfs}, and there have been significant new developments since Lattice 2022 that I will discuss further below. At hadronic energy scales, $b\to s \ell^+\ell^-$ decays are described by the weak effective Hamiltonian \cite{Grinstein:1988me}
\begin{equation}
 \mathcal{H}_{\rm eff}\:\: =\:\: -\frac{4 G_F}{\sqrt{2}}V_{tb}V_{ts}^* \sum_i C_i {O_i},
\end{equation}
where the most important operators are
\begin{eqnarray}
\nonumber {O_1} \!\!&=&\!\! \bar{c}^b\gamma^\mu b_L^a\:\:\:\bar{s}^a\gamma_\mu c_L^b, \phantom{\big]} \\
\nonumber {O_2} \!\!&=&\!\! \bar{c}^a\gamma^\mu b_L^a\:\:\:\bar{s}^b\gamma_\mu c_L^b, \phantom{\big]} \\
\nonumber {O_7}    \!\!&=&\!\! (e\: m_b)/(16\pi^2)\:\: \bar{s} \sigma^{\mu\nu} b_R \:\:\: F_{\mu\nu}^{(\rm e.m.)}, \phantom{\big]} \\
\nonumber {O_{9\ell}}    \!\!&=&\!\! e^2/(16\pi^2)\:\: \bar{s} \gamma^\mu b_L\:\:\: \bar{\ell} \gamma_\mu \ell, \phantom{\big]} \\
{O_{10\ell}} \!\!&=&\!\! e^2/(16\pi^2)\:\: \bar{s} \gamma^\mu b_L\:\:\: \bar{\ell} \gamma_\mu \gamma_5 \ell.
\end{eqnarray}
The SM values of the Wilson coefficients, evaluated using EOS \cite{EOSAuthors:2021xpv} in the $\overline{\text{MS}}$ scheme and at $\mu=4.2$ GeV, are $C_1\approx-0.288$, $C_2\approx1.010$, $C_7\approx-0.336$, $C_{9\ell}\approx 4.275$, $C_{10\ell}\approx-4.160$ (independent of the lepton flavor $\ell$).

\begin{figure}
\begin{center}
\includegraphics[width=0.25\linewidth]{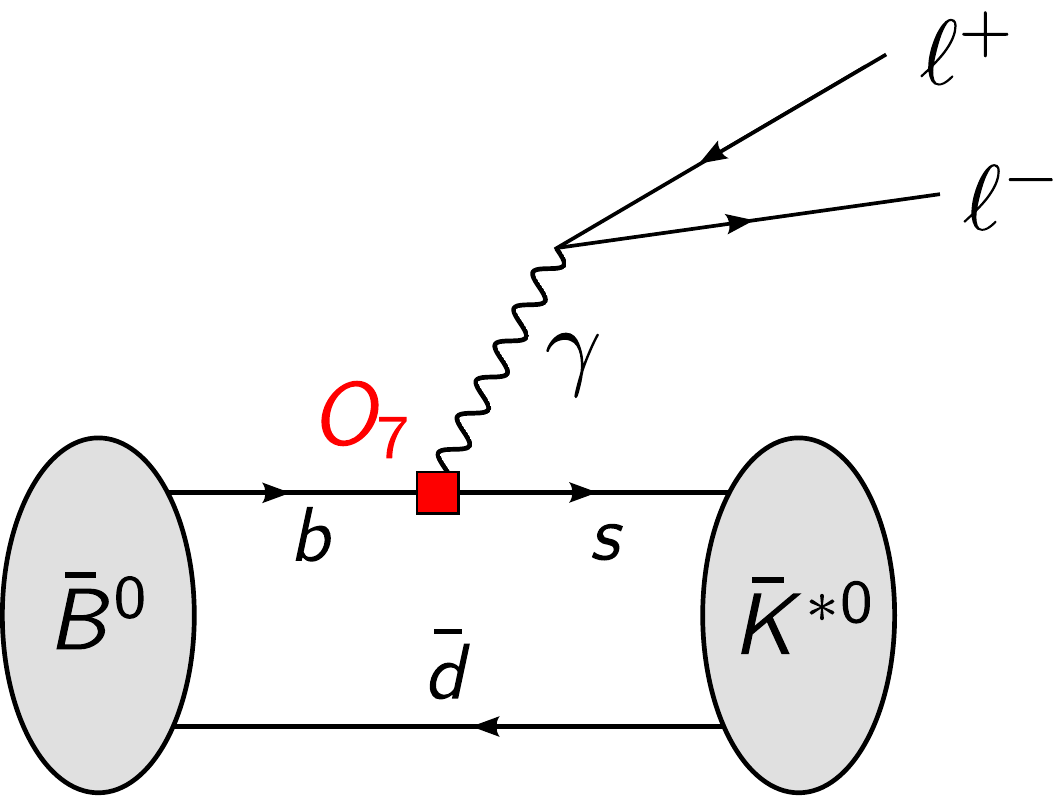} \hspace{2ex} \includegraphics[width=0.275\linewidth]{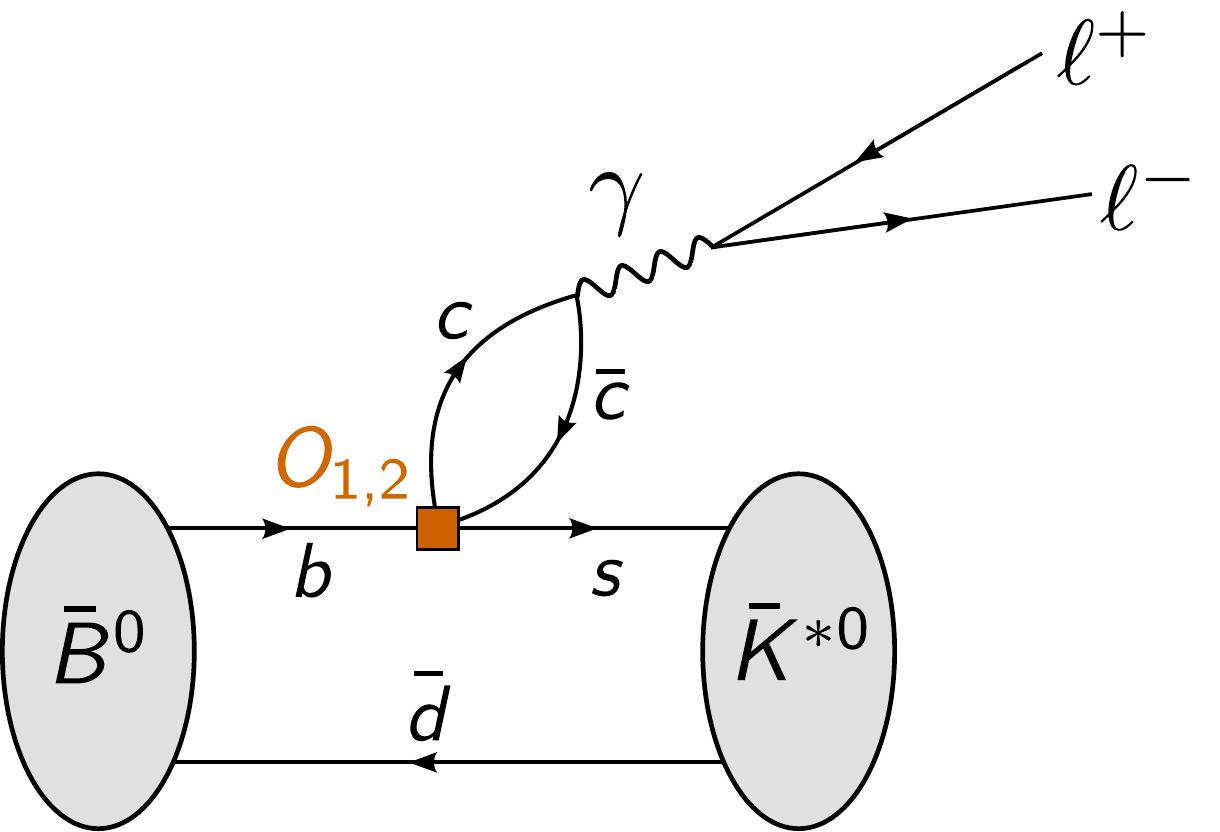} \hspace{0.5ex} \includegraphics[width=0.25\linewidth]{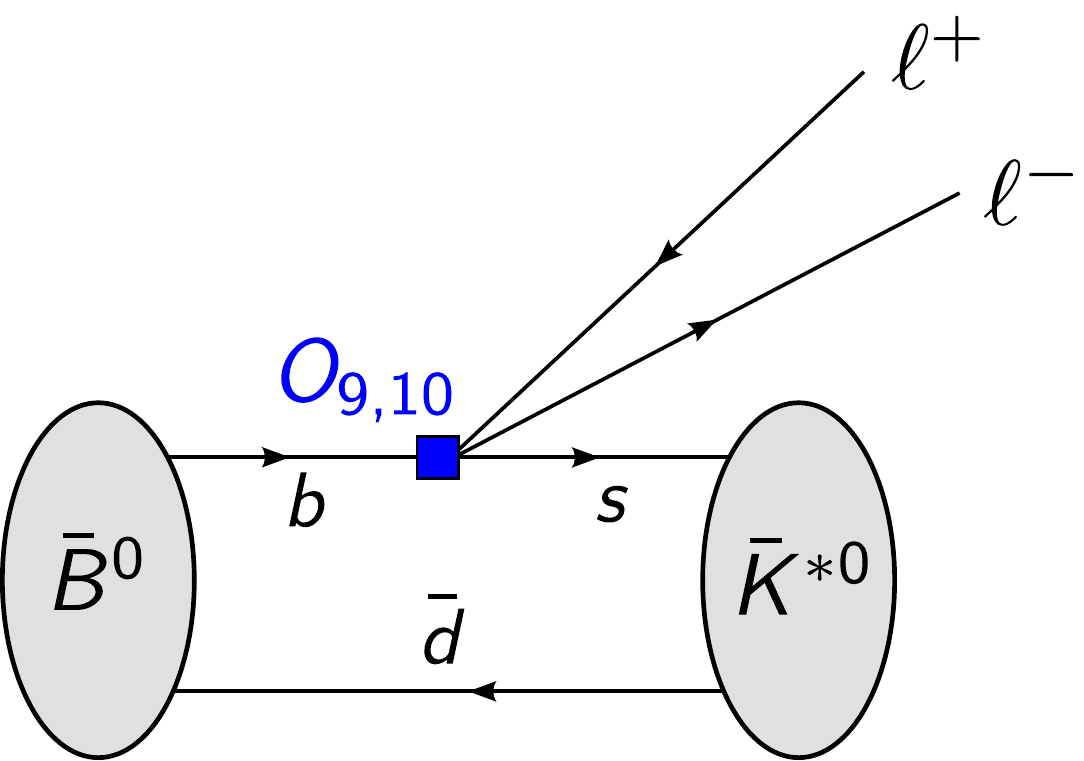}
\end{center}
\caption{\label{eq:bslldiagrams}Diagrams illustrating the contributions from different weak effective operators to the $\bar{B}^{0} \to \bar{K}^{*0} \ell^+ \ell^-$ amplitude.}
\end{figure}

\begin{figure}
 \includegraphics[width=0.49\linewidth]{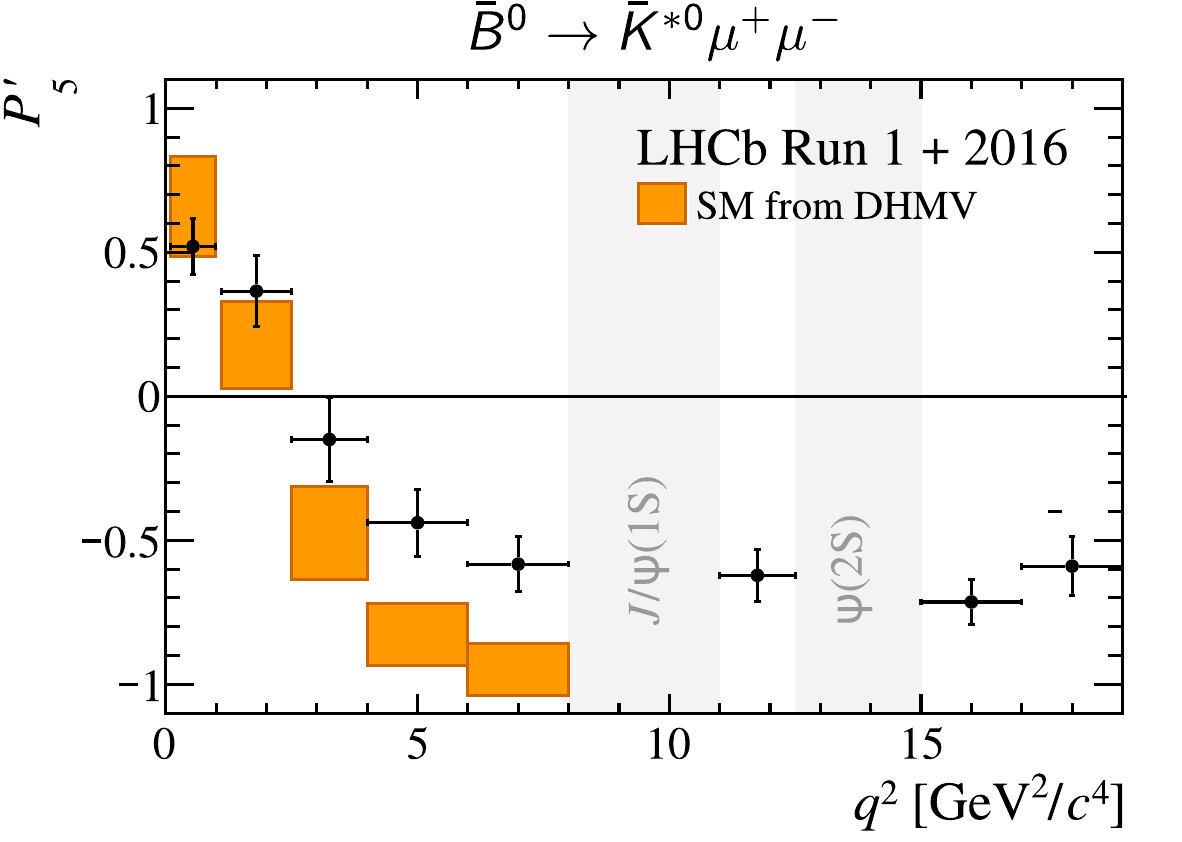} \hfill \includegraphics[width=0.49\linewidth]{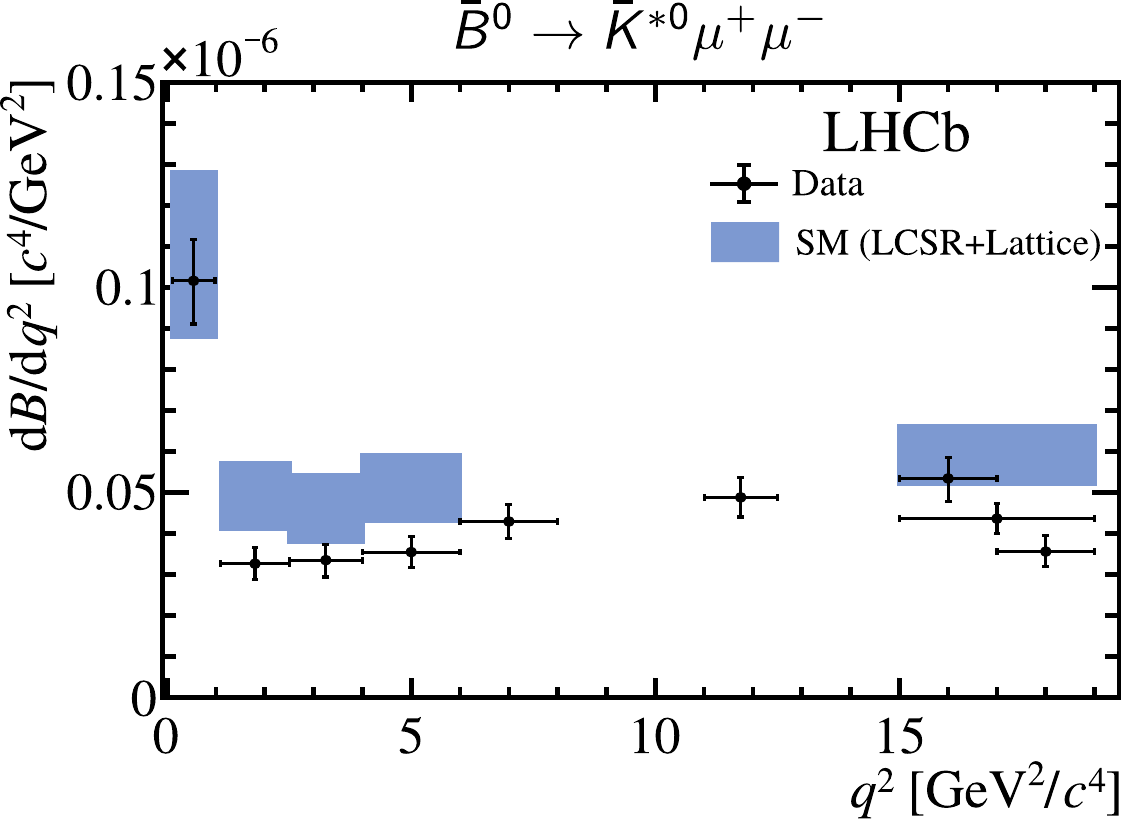} 
 
 \includegraphics[width=0.49\linewidth]{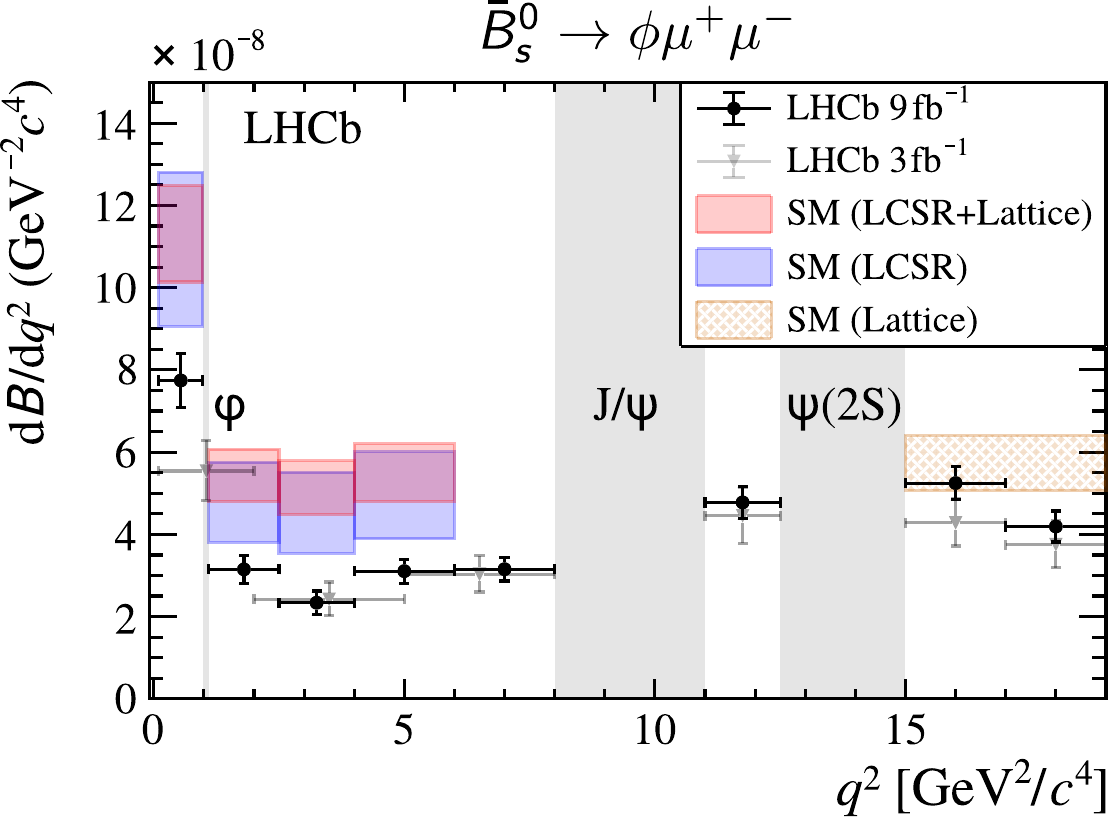} \hfill \includegraphics[width=0.49\linewidth]{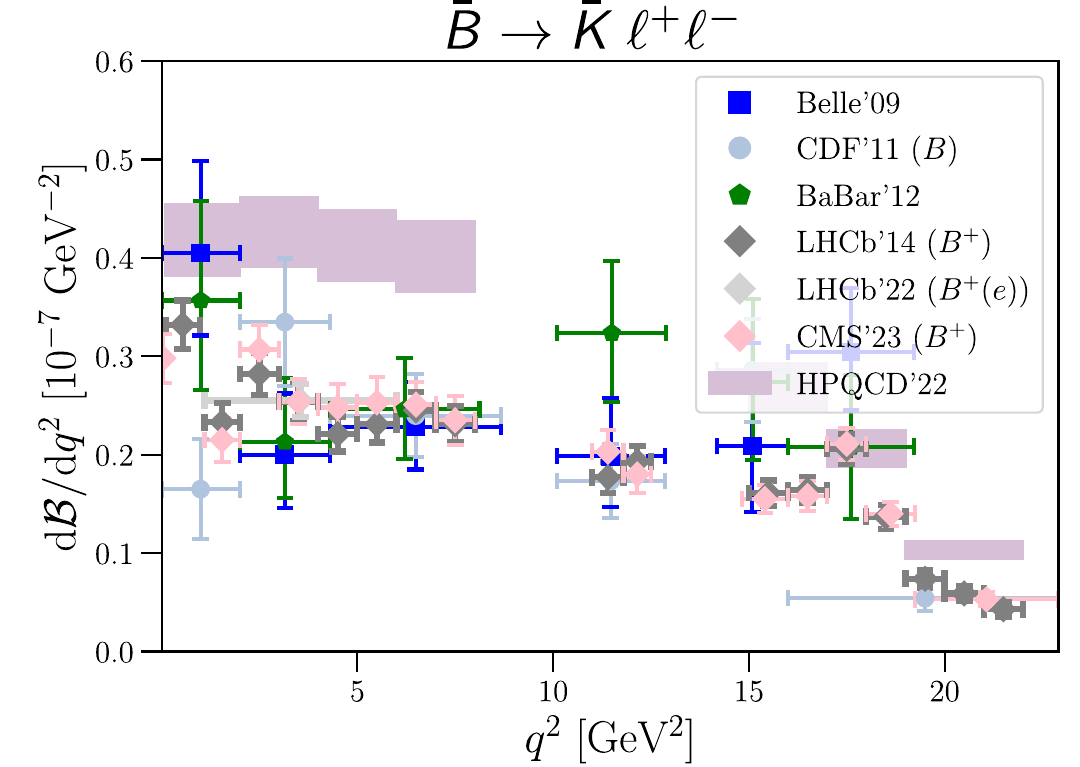}

 \vspace{2ex}
 
 \includegraphics[width=0.49\linewidth]{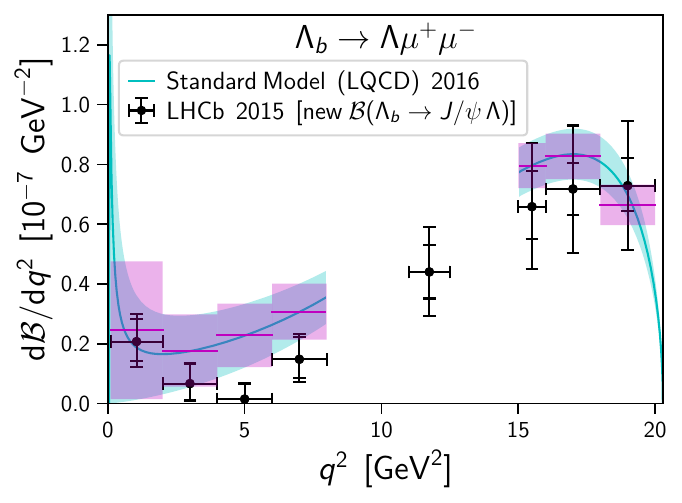} \hfill \includegraphics[width=0.49\linewidth]{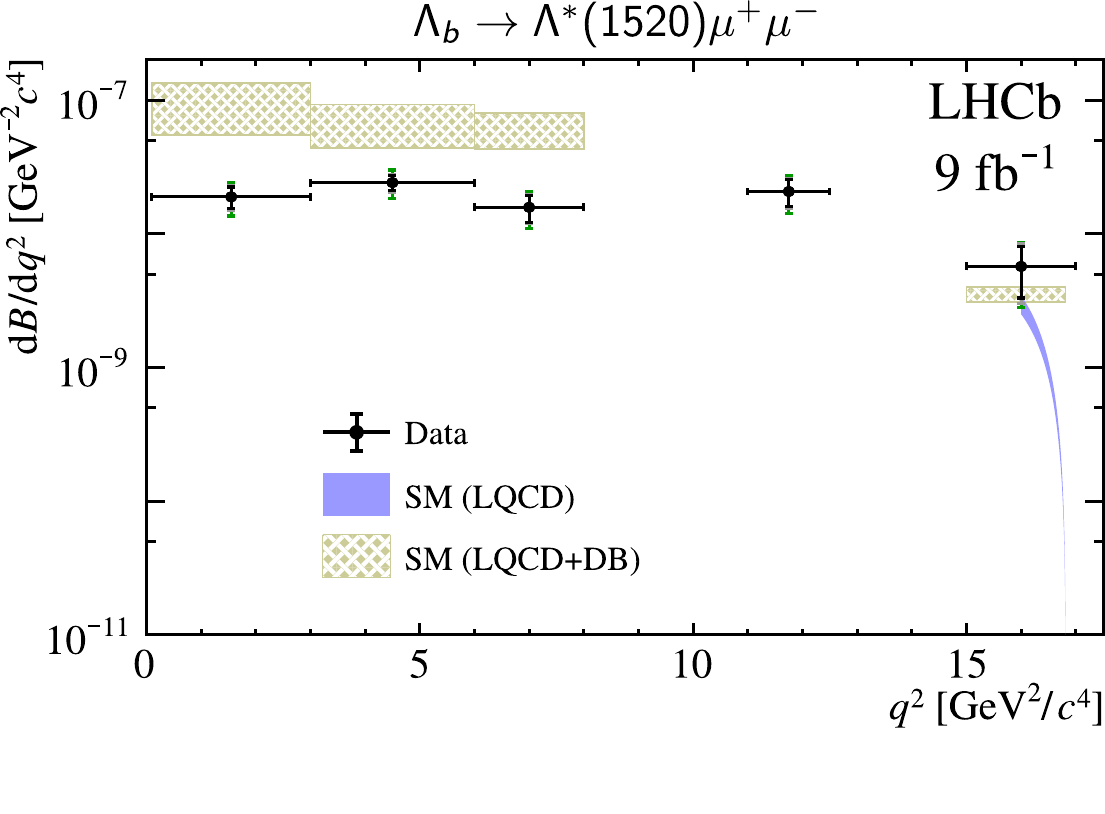}
 
 \caption{\label{fig:bssresults}Selected experimental results for $b\to s\ell^+\ell^-$ decays compared to SM predictions: the $\bar{B}^{0} \to \bar{K}^{*0} \mu^+ \mu^-$ angular observable $P_5^\prime$ \cite{LHCb:2020lmf} (SM prediction from Refs.~\cite{Descotes-Genon:2014uoa,Khodjamirian:2010vf}), the $\bar{B}^{0} \to \bar{K}^{*0} \mu^+ \mu^-$ differential branching fraction \cite{LHCb:2016ykl} (SM prediction from Refs.~\cite{Bharucha:2015bzk,Horgan:2013hoa,Horgan:2015vla}), the $\bar{B}_s^{0} \to \phi \mu^+ \mu^-$ differential branching fraction \cite{LHCb:2021zwz} (SM predictions from Refs.~\cite{Altmannshofer:2014rta,Bharucha:2015bzk,Straub:2018kue,Horgan:2013pva,Horgan:2015vla}), the $\bar{B} \to \bar{K} \ell^+ \ell^-$ differential branching fractions for both $\ell=\mu$ and $\ell=e$ (plot by Patrick Koppenburg with experimental results from Refs.~\cite{Belle:2009zue,CDF:2011grz,BaBar:2012mrf,LHCb:2014cxe,LHCb:2022vje,CMS-PAS-BPH-22-005} and SM prediction from Ref.~\cite{Parrott:2022zte}), the $\Lambda_b\to\Lambda\mu^+\mu^-$ differential branching fraction (my plot with LHCb data from \cite{LHCb:2015tgy}, with an updated normalization as discussed in Ref.~\cite{Blake:2019guk} and SM prediction from \cite{Detmold:2016pkz}), and the $\Lambda_b\to\Lambda^*(1520)\mu^+\mu^-$ differential branching fraction \cite{LHCb:2023ptw} (SM predictions from Refs.~\cite{Meinel:2021mdj,Amhis:2022vcd}; I removed the quark-model predictions from the figure).}
\end{figure}

The way in which the above operators contribute to, for example, the $\bar{B}^{0} \to \bar{K}^{*0} \ell^+ \ell^-$ decay amplitude is illustrated in Fig.~\ref{eq:bslldiagrams}. For a generic decay $H_b \to H_s \ell^+\ell^-$, evaluating the contributions from $O_7$, $O_9$, $O_{10}$ requires the hadronic matrix elements $\langle H_s(p^\prime) | \: \bar{s} \Gamma b \: | H_b(p) \rangle$ that correspond to the local form factors and have been calculated using lattice QCD for several different combinations of $H_b$ and $H_s$ hadrons \cite{Bouchard:2013eph,Horgan:2013hoa,Horgan:2015vla,Bailey:2015dka,Detmold:2016pkz,Meinel:2021mdj,Parrott:2022rgu}. Note that some of the $H_s$ considered are unstable under the strong interactions [$K^*$, $\phi$, $\Lambda^*(1520)$], and the calculations published to date neglected this property. Going beyond this approximation with the proper Lellouch-L\"uscher approach \cite{Briceno:2014uqa} is needed and feasible for the $B\to K^*(\to K\pi)$ form factors \cite{Leskovec:2024pzb}. For the $B_s\to \phi$ form factors with the much narrower $\phi$, I think it is worth doing new calculations even with the simple single-hadron treatment.

The operators $O_{1,...,6}$, $O_8$ contribute through matrix elements with an additional nonlocal insertion of the quark electromagnetic current: in Minkowski spacetime,
\begin{equation}
\int \mathrm{d}^4 x \:\:e^{iq\cdot x}\,\langle H_s(p^\prime) | \: \mathsf{T} \:  O_i(0) \: J^\mu_{\rm e.m.}(x) \: | H_b(p) \rangle. \label{eq:bnonlocal}
\end{equation}
Computing these matrix elements is very challenging for Euclidean lattice QCD (see Ref.~\cite{Nakayama:2020hhu} for first steps), and they are currently instead being approximated in the continuum using a local operator-product expansion (OPE) at high $q^2$ \cite{Grinstein:2004vb,Beylich:2011aq} and QCD factorization or a light-cone OPE at low $q^2$ \cite{Beneke:2001at,Khodjamirian:2012rm}. This usually means staying away from the region with $q^2\sim m_{J/\psi,\psi'}^2$, where the matrix elements of $O_{1,2}$ are enhanced dramatically though the charmonium resonances, but recent work using dispersion relations has also taken advantage of the experimental measurements of $\mathcal{B}(H_b\to H_s\,J/\psi^{(\prime)})$ \cite{Gubernari:2020eft,Gubernari:2022hxn}.

There has been substantial excitement about $b\to s\ell^+\ell^-$ decays since 2013, when deviations between measurements and SM predictions were first seen in $\bar{B}^{0} \to \bar{K}^{*0} \mu^+ \mu^-$ at both low $q^2$ \cite{Descotes-Genon:2013wba} and high $q^2$ \cite{Horgan:2013pva} (also for $\bar{B}_s^0 \to \phi \mu^+ \mu^-$ \cite{Horgan:2013pva} and $\bar{B}\to\bar{K}\mu^+\mu^-$ \cite{Du:2015tda}) that can be explained by a negative shift in the Wilson coefficient $C_{9\mu}$. These deviations, with a general trend of branching fractions below the SM predictions, persist to date, as shown in Fig.~\ref{fig:bssresults}. Moreover, LHCb measurements \cite{LHCb:2014vgu,LHCb:2017avl,LHCb:2019hip,LHCb:2019efc,LHCb:2021trn,LHCb:2021lvy} of muon-versus-electron ratios such as $R_K \:\equiv\: \int_{1\:{\rm GeV}^2}^{6\:{\rm GeV}^2} \frac{\mathrm{d}\mathcal{B} (B^+ \to K^+ {\mu^+\mu^-})}{\mathrm{d}q^2}\mathrm{d}q^2/\int_{1\:{\rm GeV}^2}^{6\:{\rm GeV}^2} \frac{\mathrm{d}\mathcal{B} (B^+ \to K^+ {e^+ e^-})}{\mathrm{d}q^2}\mathrm{d}q^2$ gave values below the theoretically clean SM predictions equal to $1+\mathcal{O}(10^{-3})$, consistent with the scenario $C_{9\mu}<C_{9}^{\rm SM}$ and $C_{9e}=C_{9}^{\rm SM}$. The big news from December 2022 \cite{LHCb:2022vje} is that the LHCb results for these ratios had an uncontrolled systematic error (hadrons misidentified as electrons), and the corrected LHCb analysis gives $R_{K^{(*)}}$ consistent with unity. It is important to note, however, that $\frac{\mathrm{d}\mathcal{B} (B \to K^{(*)} {\mu^+ \mu^-})}{\mathrm{d}q^2}$ is unchanged and $\frac{\mathrm{d}\mathcal{B} (B \to K^{(*)} {e^+ e^-})}{\mathrm{d}q^2}$ has moved lower, \emph{farther away} from SM predictions. As a result, a new-physics scenario with both $C_{9\mu}<C_{9}^{\rm SM}$ and $C_{9e}<C_{9}^{\rm SM}$ is now favored by global fits (see, e.g., Ref.~\cite{Alguero:2023jeh}); viable models with this effect are discussed, for example, in Ref.~\cite{Greljo:2022jac}. These findings critically depend on QCD calculations, and improving these calculations is therefore now more important than ever.

On that note, at this conference, progress with new lattice-QCD calculations of $b\to s$ form factors was reported by Hwancheol Jeong for $B\to K$, using using HISQ light quarks and Fermilab $b$ quarks \cite{Jeong}, and by me for $\Lambda_b \to \Lambda$, using domain-wall light quarks and RHQ $b$ quarks \cite{Meinel:2023wyg}.

\FloatBarrier
\subsection{Rare kaon and hyperon decays}

Rare decays of \emph{strange} hadrons also provide powerful constraints on physics beyond the Standard Model \cite{Goudzovski:2022scl}. For some decay modes, theory predictions are far more precise than current experimental results, while for others, the situation is the opposite. Theoretically cleanest are the dineutrino modes $K_L \to \pi^0 \bar{\nu}\nu$ and $K^+ \to \pi^+ \bar{\nu}\nu$ \cite{Buras:2015qea,Buras:2015yca}, for which the QCD uncertainties of the SM predictions of the decay rates are 1.5\% and 4\% respectively \cite{Goudzovski:2022scl}, while the current experimental results are \cite{KOTO:2018dsc}
\begin{equation}
\mathcal{B}(K_L \to \pi^0 \bar{\nu}\nu)< 3.0\times 10^{-9}\:\:(90\%\:\:{\rm CL})
\end{equation}
and \cite{NA62:2021zjw}
\begin{equation}
\mathcal{B}(K^+ \to \pi^+ \bar{\nu}\nu) =  (10.6_{-3.4}^{+4.0}|_{\rm stat} \pm 0.9|_{\rm syst})\times 10^{-11}.
\end{equation}
The 4\% QCD uncertainty in the SM prediction of $\mathcal{B}(K^+ \to \pi^+ \bar{\nu}\nu)$ is dominated by the contribution from nonlocal hadronic matrix elements, which can be calculated with lattice QCD \cite{Christ:2019dxu} and will become more relevant as the experimental precision improves in the future.

Some of the experimental results for charged-lepton modes are \cite{ParticleDataGroup:2022pth}
\begin{align}
\nonumber \mathcal{B}(K^+\to \pi^+ e^+ e^-)&=3.00(9)\times 10^{-7}, \\
\nonumber \mathcal{B}(K^+\to \pi^+ \mu^+ \mu^-)&=9.17(14)\times 10^{-8}, \\
\mathcal{B}(\Sigma^+\to p^+\mu^+\mu^-)&=2.4^{+1.7}_{-1.3}\times 10^{-8}.
\end{align}
The SM predictions for these processes are dominated by nonlocal hadronic matrix elements, which can, in principle, be calculated on the lattice \cite{RBC:2022ddw,Erben:2022igb}. The current lattice results still have uncertainties larger than the central values of the experimental measurements (see Ref.~\cite{Blum:2022wsz} for a discussion of future prospects).

\begin{figure}
\includegraphics[width=0.3\linewidth]{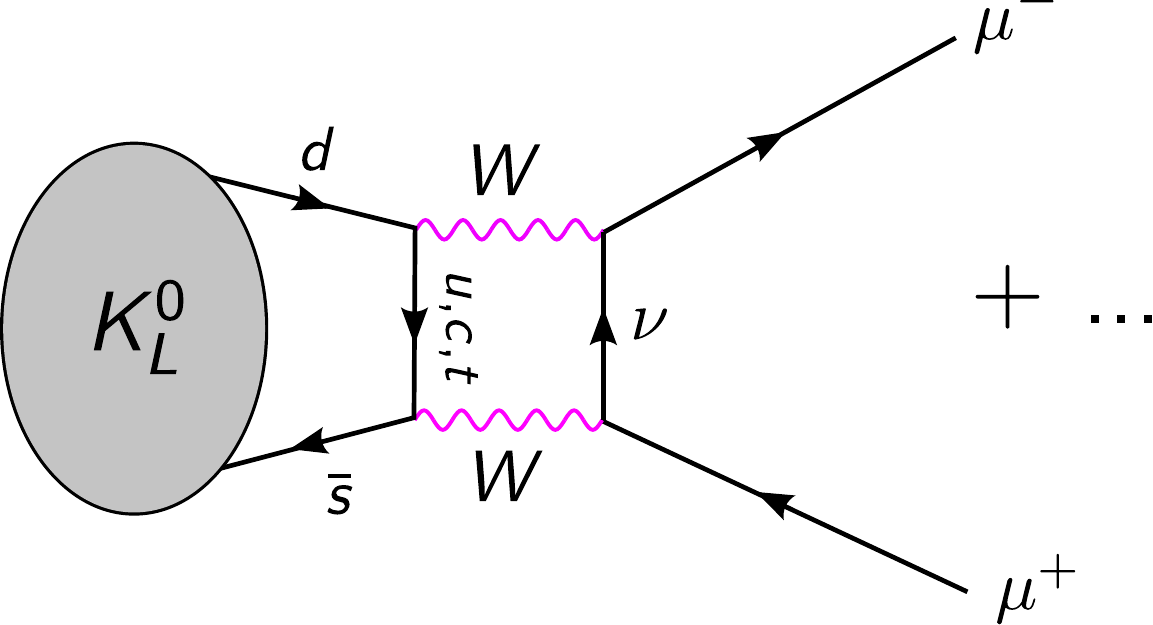} \hfill \includegraphics[width=0.5\linewidth]{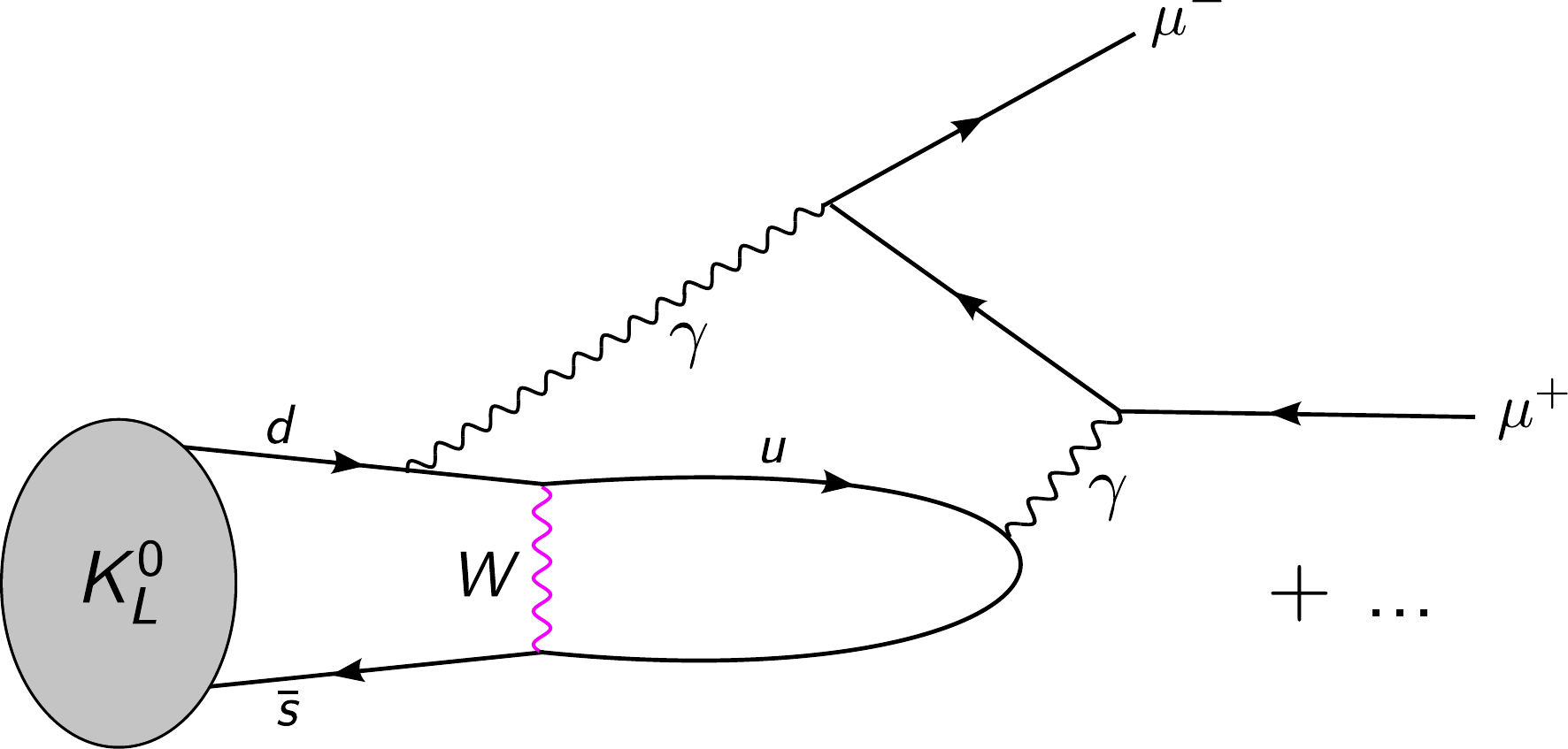}
\caption{\label{fig:KLmumu}Examples of diagrams contributing to $K_L \to \mu^+ \mu^-$ in the Standard Model.}
\end{figure}

Another interesting rare kaon decay is $K_L \to \mu^+ \mu^-$. Here, diagrams with exchanges of two weak bosons as well as diagrams with one weak boson and two photons (see Fig.~\ref{fig:KLmumu}) are important. The contribution of the former has a 5\% QCD uncertainty, mainly from the diagrams with charm quarks \cite{Gorbahn:2006bm}. The absorptive part of the two-photon contribution, which nearly saturates the SM prediction of $\mathcal{B}(K_L \to \mu^+ \mu^-)$, can be obtained with high precision from the measured $\mathcal{B}(K_L\to\gamma\gamma)$, but the remaining dispersive part of the two-photon contribution is of comparable magnitude to the two-weak-boson contribution \cite{Gorbahn:2006bm} and, while very challenging, is a possible target for lattice-QCD calculations \cite{Christ:2020bzb}. A first step in this direction, a lattice-QCD calculation of $\pi^0\to e^+e^-$, was recently completed \cite{Christ:2022rho}.

At this conference, En-Hung Chao discussed a framework to compute the two-photon contribution to $K_L \to \mu^+ \mu^-$ (which requires identifying and subtracting unwanted contributions that appear with Euclidean time) and presented preliminary numerical results for the quark-connected diagrams \cite{Chao:2023cxp}. Bai-Long Hoid reviewed the application of the continuum, dispersive approach to $\pi^0\to e^+e^-$ and $K_L \to \mu^+ \mu^-$ and its relation to lattice QCD \cite{Hoid:2023has}. Raoul Hodgson gave an update \cite{Hodgson} on the exploratory lattice calculation \cite{Erben:2022igb} of the rare hyperon decay $\Sigma^+ \to p \ell^+ \ell^-$. Amarjit Soni presented an overview of rare kaon decays and emphasized the role of $K^0\to \pi^0\ell^+\ell^-$ \cite{Schacht:2023vsz}.

\FloatBarrier
\subsection{CP violation in charm decays}

\noindent CP violation in charm decays was discovered in 2019 by LHCb, with the time-averaged result \cite{LHCb:2019hro}
\begin{equation}
 \Delta A_{CP} = A_{CP}(K^+K^-)-A_{CP}(\pi^+\pi^-) = (-15.4 \pm 2.9 )\times 10^{-4}, \label{eq:charmCPexp}
\end{equation}
where
\begin{equation}
 A_{CP}(f; t) = \frac{\Gamma(D^0(t) \to f)-\Gamma(\overline{D}^0(t) \to f)}{\Gamma(D^0(t) \to f) + \Gamma(\overline{D}^0(t) \to f)}.
\end{equation}
The time-dependent analysis shows that $\Delta A_{CP}$ is dominated by direct CP violation. More recently, LHCb also determined the individual asymmetries \cite{LHCb:2022lry}. It is an open question whether the result (\ref{eq:charmCPexp}) is compatible with the Standard Model or a signal of new physics, because theoretical predictions for $\Delta A_{CP}$ vary substantially depending on the methods used to estimate the nonperturbative QCD contributions. For example,
\begin{align}
\nonumber & \Delta A_{CP}^{\rm SM}\approx 2\times 10^{-4} \hspace{2ex}\text{\cite{Khodjamirian:2017zdu}}, \\
\nonumber & \Delta A_{CP}^{\rm SM}\approx -4\times 10^{-4} \hspace{2ex}\text{\cite{Pich:2023kim}}, \\
& \Delta A_{CP}^{\rm SM}\approx -16\times 10^{-4} \hspace{2ex}\text{\cite{Schacht:2021jaz}}.
\end{align}
A lattice-QCD calculation of the relevant $D\to\pi\pi$ and $D\to K\bar{K}$ amplitudes is therefore an important long-term goal. This is even more challenging than $K\to\pi\pi$ due to the several coupled channels (including $\pi\pi\pi\pi$) that need to be considered to understand the finite-volume final states in the $D$-meson energy region for physical quark masses. A necessary first step was the generalization of the Lellouch-L\"uscher approach to coupled two-body channels \cite{Hansen:2012tf}.

At this conference, Maxwell Hansen presented a pilot lattice study of hadronic $D$ decays using stabilized Wilson fermions \cite{Hansen}. This work considers $D\to K\pi$ at the $SU(3)$ flavor-symmetric point, which avoids power-divergent operator mixing otherwise present with Wilson fermions and allows the use of the single-channel Lellouch-L\"uscher factor. He showed preliminary results for the renormalization of the four-quark operators and for the finite-volume spectra and the corresponding scattering phase shifts.

\FloatBarrier
\section{Conclusions}

\noindent Quark flavor physics is exciting and may lead to the discovery of physics beyond the Standard Model. We already see interesting deviations between measurements and SM predictions that have inspired substantial model-building work and demonstrate possible routes to discovery.
Lattice-QCD calculations are essential for quark flavor physics. There has been excellent progress, and we need to continue and expand this work to make the best use of existing precise measurements and to keep up with the expected experimental progress (see, e.g., Refs.~\cite{DiCanto:2022icc,Goudzovski:2022scl}) in the coming years.
It is very valuable to have multiple calculations from different groups with different methods. Tensions between some of the lattice results for semileptonic form factors have emerged, indicating that uncertainties were underestimated in some cases. Thankfully, many new calculations with improved methods are already underway.

\FloatBarrier

\providecommand{\href}[2]{#2}\begingroup\raggedright\endgroup

\end{document}